\documentclass[a4paper,11pt,twoside]{scrbook}
\usepackage[hmargin=3cm,tmargin=4.5cm,bmargin=4.5cm]{geometry}	% Custom margins for b5-conversion
\usepackage{amsmath}		% Better math formatting.
\usepackage{amssymb}		% Better math formatting.
\usepackage{bm}		        % Better math formatting.
\usepackage{wrapfig}
\usepackage{mathrsfs}		% Raph Smith's Formal Script support.
\usepackage{graphicx}		% Better graphics support.
\usepackage{multirow}		% Allows multilrow/column table/array cells.
\usepackage[labelfont=bf,font=small,font=sf]{caption}		% Better captions
\usepackage{fancyhdr}		% Fancy headers
\usepackage{pdfpages}		% Include pdf-documents.
\usepackage{float}
\usepackage[T1]{fontenc}
\usepackage{pdfpages}
\usepackage{cite}

\usepackage{yhmath}
\usepackage{lipsum}			% Automatically generates blocks of placeholder text.
\usepackage{color}			% Allows colored text.
\definecolor{dkgreen}{rgb}{0,0.4,0}

\fancypagestyle{plain}{ %
	\fancyhf{} % remove everything
	\fancyfoot[C]{\sf\textbf{\thepage}}
	}

\restylefloat{figure}

\newcommand{\beq}{\begin{equation}}
\newcommand{\eeq}{\end{equation}}

\newcommand{\bqa}{\begin{eqnarray}}
\newcommand{\eqa}{\end{eqnarray}}

\author{Amna Noreen}
\title{Very-High-Precision Calculations in Physics}
\date{November 20, 2012}

\begin{document}

% THE FIRST STUFF: TITLE PAGE,
% ACKNOWLEDGEMENTS, TABLE OF CONTENTS,
% ETC.
%
\maketitle

\frontmatter
	% Preamble
%	\input{phd_template_preamble}

%\begin{figure}[H]
%\begin{center}

%\includegraphics[clip, trim= .1ex .1ex .1ex .1ex, width=0.95\textwidth]{Dua.jpg}
%\end{center}
%\end{figure}
%\vspace{-15ex}
%\begin{center}
%\textit{\Huge Oh My Lord! increase me in my Knowledge.}\\
%(Holy Quran, Surah Ta Ha, 20:114.)
%\end{center}
\chapter*{}

\vspace{22ex}

\begin{center}
{\huge
\emph{O my Lord! Increase me further in knowledge.}
}
\\[5ex]
{\large
(Holy Quran, Surah Ta Ha, 20:114.)
%\emph{I dedicate this work to the beloved parents of both of us}
}
\end{center}

\chapter*{}

\vspace{30ex}

{\large
\begin{center}

\emph{I dedicate this work and give special thanks to my lovely husband,
Asif Mushtaq,\\
for being there for me throughout the entire doctorate program.}\\[5ex]

\emph{I dedicate this work to the beloved parents of both of us}

\end{center}
}

\chapter*{Preface}

This thesis is submitted by the author as part of the requirements for the
degree Philosophiae Doctor at the Norwegian University of Science and
Technology (NTNU). It is the conclusion of a little more than four years of work.\\

\noindent
My supervisor has been Professor K{\aa}re Olaussen. The work has been
performed at the Norwegian University of Science and Technology.

\chapter*{Acknowledgement}% This section is not numbered

\begin{center}
\textbf{Thanks to ALMIGHTY ALLAH for everything.}\\
\end{center}

I would like to express my sincere gratitude to my supervisor
Professor~Dr.~K{\aa}re Olaussen, for the continuous support
of my Ph.D study and research, for his patience, motivation,
enthusiasm and immense knowledge. He has made our
survival in Norway possible.
I can not imagine having a better advisor and mentor for my Ph.D studies.
I would like to express my thanks to his wife Aud for her motherly attitude.
I would like to thank my co-supervisor Associate Professor Ingjald {\O}verb{\o}
for his encouragement and discussions.

I acknowledge Higher Education Commision HEC of Pakistan
and Department of Physics for their financial support.

I thank all friends and colleagues I have interacted
with in the Department of Physics at NTNU during these years.
I thank all office secretaries for teaching me Norwegian
in a friendly atmosphere. 

At this juncture I think of my parents whose selfless sacrificial life
and great efforts and unceasing prayers have 
enabled me to reach the present position in life. I am forever indebted to my 
mother who planted the seed of idea that I should become a doctor,
and her encouragement in attaining this goal (even though I have aimed for a
different kind of doctor). I would like to express my deepest gratitude to my
parents-in-law for their love, prayers and
permission to undertake these higher studies. 
I am grateful to my sister Saadia
for her love and support during the last year.

My special thanks to my husband Asif Mushtaq for his constant support,
encouragement, pleasant association, patience and help me in various forms.
I can never forget that he sacrificed his permanent university job only
because of his family; this was a painful experience for us.
At this time I must not forget my loving daughters Zemal and Zoya for their 
everlasting love. I can never forget difficult time at the beginning of my Ph.D study,
when I  came Norway alone. At that time my elder daughter was 1 year and six
months old and my mother-in-law took the resposibility for her.
And my younger daughter was only 3 months old and my father took
the responsiblity for her. My special thanks to my mother-in-law and my father.

Finally I thank all those who have helped me directly or indirectly
towards the completion of my thesis. Anyone missed in this 
acknowledgement is also thanked.
%\end{document}

%\input{Articles}

\tableofcontents

\mainmatter

%
% THE MAIN STUFF
%
% \Huge{\textbf{List of Articles}}\\
% \Large{\textbf{Article I}}\\
% \Large{\textbf{Article II}}\\
% \Large{\textbf{Article III}}\\
% \Large{\textbf{Article IV(a)}}\\
% \Large{\textbf{Article IV(b)}}\\
% \Large{\textbf{Article V}}

\part{Introduction}

% PDF TEMPLATE: CHAPTER 1

\chapter{Physical background}

\section{What is the most accurate number in physics? \label{MostAccurate}}

What is the most accurately known number in physics? It can be argued
that theoretically this is the ratio between the perimeter and diameter of
a circular disk, also known as $\pi$, quite recently computed to about $10^{13}$
decimals \cite{ShigeruKondo}. This precision is not
quite matched by experimental observations. It can also be criticized
for not taking into account the discrete and quantum nature of matter,
or --- at this accuracy --- even the tiny non-euclidean nature
of surrounding space.

Experimentally the best accuracy is probably what can be obtained by use of
optical frequency combs \cite{NobelPrize2005}, currently with a relative accuracy
of a few parts in $10^{17}$ (i.e.~17 decimals). 
This is expected to improve by a few orders of magnitude
during the next decades, cf.~Fig~2 of reference \citen{NobelPrize2005}.
Hence, there are situations where it makes sense to compute physical quantities to 20 decimals
precision or better --- provided that the physical model is known accurately enough.

The latter is usually not the case. The physical quantity where experimental 
and theoretical values are in best agreement is probably the magnetic moment
of the electron. This quantity is measured to about 13 decimals  \cite{PDG},
with the computation of the theoretical contributions from Quantum Electrodynamics
(QED) recently completed to fifth order in the fine structure constant $\alpha$
\cite{Electron_g-2_10thOrder}. This leads to agreement between experiment
and theory to about 13 decimals without adjustable parameters.
Also the QED contributions to the muon magnetic moment has been computed
to fifth order in $\alpha$ \cite{Muon_g-2_10thOrder}, but for this quantity
the theoretical contributions from other sources (like hadronic contributions to vacuum
polarization) are larger, and the experimental uncertainty is also larger.

\section{The role of ordinary differential equations in physics}

Most of physics can be described locally in space and time, hence mathematically
by differential equations. Differential equations therefore form a central part of
theoretical physics, both on the elementary and advanced level.
In most cases the relevant equations are \emph{partial}.
But by symmetry reductions, or more systematically for linear equations
by the method of separation of variables, they can be reduced to ordinary
differential equations. This greatly increases the prospects of finding
solutions, and of understanding the properties of such solutions.

There is a powerful and quite complete method of solving ordinary
linear homogeneous equations, starting with works by Fuchs \cite{Fuchs} and
Frobenius \cite{FrobeniusMethod}. In this method the solution is expanded in a 
convergent (generalized) power series, a Frobenius series, around ordinary or
regular singular points of the equation in the complex plane.
This method can be combined with analytic continuation to extend the solution
beyond the radius of convergence of each power series.

\subsection{Schr{\"o}dinger equation in one dimension}

The Schr{\"o}dinger equation is a partial differential equation that governs the
time evolution of quantum mechanical wave-functions $\Psi(\bm{q},t)$, where $\bm{q}$
denotes the position(s) of the particle(s).
It gives a good description of the quantum motion of non-relativistic particles interacting
instantaneously with each other.
%The wavefunction can be found by solving the differential equation for a particular problem.\\
In the one-dimensional single-particle case it reads
\begin{equation}
  \left[-\frac{\hbar^{2}}{2m}\frac{\partial^2 }{\partial q^2} + V(q)\right]
  \Psi(q,t) =  i \hbar\frac{\partial \Psi(q,t)}{\partial t}.
  \label{Schrodingerequation}
\end{equation}
This equation can be reduced to an ordinary differential equation by
separation of variables. Assuming a solution in product form,
\begin{equation*}
  \Psi(q,t) = \psi(q)\,T(t).
\end{equation*}
and substituting into equation~(\ref{Schrodingerequation}), one finds $T(t) = \text{e}^{-i{E}t/{\hbar}}$,
and
\begin{equation*}
  \left[-\frac{\hbar^{2}}{2m}\frac{\partial^2 }{\partial q^2} + V(q)\right]\psi(q) 
  = E\;\psi(q).
\end{equation*}
%\indent
We assume the potential to be a low-order even polynomial,
\begin{equation}
 V(q) = \sum_{n=0}^N \nu_n q^{2n}.
\end{equation}
For numerical computations one must use dimensionless quantities. We
first introduce a dimensionless length $x = q/\lambda$ such that
the Schr{\"o}dinger equation becomes
\begin{equation}
  \left[-\frac{\partial^2 }{\partial x^2} + x^{2N} + \sum_{n=0}^ {N-1} v_{n}\, x^{2n}\right]
  \psi(x) =  \varepsilon\, \psi(x),
  \label{1dSchrodingerEquation}
\end{equation}
with $\lambda^{2N+2} = \hbar^2/2m V_N$, $v_{n} = 2m V_n \lambda^{2n+2}/\hbar^2$, and
$\varepsilon = {2mE \lambda^2}/{\hbar^{2}}$. Here the scaling coefficient $\lambda$
has been chosen to give unit coefficients in front of the $\partial^2/\partial x^2$ and the $x^{2N}$ terms.
Other choices may sometimes be more convenient, in particular it is natural to generalize to
the case where
\begin{equation*}
  -\frac{\partial^2}{\partial x^2} \rightarrow -s^2 \frac{\partial^2}{\partial x^2},
\end{equation*}
with $s^2$ usually a small number.
It is common to think of it as $\hbar^2/2 m$, but since this quantity is
not dimensionless it is not a true small parameter of the equation.

The solutions of equation~(\ref{1dSchrodingerEquation}) be expanded
in a power series in $x^2$. The radius of convergence of this power
series is infinite. One may make an analytic continuation to expand
the solution around another point $x_0$, but this will destroy the
explicit parity symmetry of the problem. The latter leads to a
doubling of expansion coefficients in the power series,
and a significant increase in the number of coefficients describing
the polynomial potential. The advantage for a numerical evaluation
is that the power series may converge faster, and
with a smaller loss of precision due to roundoff errors.

\subsection{Schr{\"o}dinger equation in higher dimensions}

The real world is not one-dimensional, but can in many situations be
treated as quite symmetric. F.i., the Schr{\"o}dinger equation for
a $D$-dimensional rotationally symmetric system,
\begin{equation}
    \left[ -\frac{\hbar^2}{2 m} \bm{\nabla}^2 + V(q)\right] \Psi(\bm{q},t) = 
    \text{i}\hbar \frac{\partial}{\partial t}\Psi(\bm{q},t),
\end{equation}
allows for a separation of variables, 
\begin{equation}
  \Psi(\bm{q},t)=\psi(q)\,{\cal Y}^{(\ell)}(\bm{\hat{q}})\,T(t).
  \label{SeparationOfVariables}
\end{equation}
Here $q\equiv \vert\bm{q}\vert$, and ${\cal Y}^{(\ell)}(\bm{\hat{q}})$
is the generalization of the spherical harmonics to $D$ dimensions. They are
independent of the length of $\bm{q}$ (i.e., scale invariant), 
hence $\bm{q}\cdot\bm{\nabla}{\cal Y}^{(\ell)}(\bm{\hat{q}}) = 0$.
The functions
\begin{equation*}
     {\cal P}^{(\ell)}(\bm{q}) \equiv q^\ell\,{\cal Y}^{(\ell)}(\bm{\hat{q}})
\end{equation*}
are homogeneous polynomials of order $\ell$ in the components of the vector $\bm{q}$,
which are also solutions of the Laplace equation
\begin{equation}
    \bm{\nabla}^2 q^{\ell}{\cal Y}^{(\ell)}(\bm{\hat{q}})  = 
    \left( \bm{\nabla}^2\,q^{\ell}\right) {\cal Y}^{(\ell)}(\bm{\hat{q}}) + 
    q^{\ell} \,\bm{\nabla}^2\,{\cal Y}^{(\ell)}(\bm{\hat{q}}) = 0. 
\end{equation}
It follows from these relations that 
\begin{align*}
    \bm{\nabla}^2\,\left[{\cal Y}^{(\ell)}(\bm{\hat{q}})\right] = 
    -\left(q^{-\ell} \,\bm{\nabla}^2 q^{\ell}\right)\,{\cal Y}^{(\ell)}(\bm{\hat{q}})
    = -\ell(\ell + D -2)\,q^{-2}\,{\cal Y}^{(\ell)}(\bm{\hat{q}}),
\end{align*}
and
\begin{align*}
    \bm{\nabla}^2\,{\cal Y}^{(\ell)}(\bm{\hat{q}})\,\psi(q) =
    {\cal Y}^{(\ell)}(\bm{\hat{q}})\,\left[\psi''(q) + (D-1) q ^{-1} \psi'(q) -\ell(\ell + D -2)\,q^{-2} \psi(q)\right].
\end{align*}
Hence, the separation of variables~(\ref{SeparationOfVariables}) leads to the radial equation
\begin{equation}
   \left[ -\frac{\hbar^2}{2m}\left(\frac{\partial^2}{\partial q^2}
     + \frac{D-1}{q}\frac{\partial}{\partial q} +\frac{\ell(\ell+D-2)}{q^2}\right) 
   + V(q) \right] \psi(q) = E\,\psi(q).
\end{equation}
This equation has a regular singular point at $q=0$, but it still has a generalized series
solution which can be found by the Frobenius method~\cite{FrobeniusMethod}.

Separations of variables can be applied to many other coordinate systems.
F.i., in three dimensions the Schr{\"o}dinger equation in zero potential can
be separated in ellipsoidal coordinates $\left(\xi_1, \xi_2, \xi_3 \right)$,
related to Cartesian coordinates by
\begin{align}
    x &= \sqrt\frac{(\xi_1^2-a^2)(\xi_2^2-a^2)(\xi_3^2-a^2)}{a^2(a^2-b^2)},\nonumber\\
    y &= \sqrt\frac{(\xi_1^2-b^2)(\xi_2^2-b^2)(\xi_3^2-b^2)}{b^2(b^2-a^2)},
    \quad \xi_1 > a > \xi_2 > b > \xi_3 > 0,\\
    z &= \frac{\xi_1 \xi_2 \xi_3}{ab},\nonumber
\end{align} 
plus 10 degenerate forms of these coordinates~\cite{MorseFeshbach}. 
The separated equations have five regular singular points, at $\pm a$, $\pm b$, and $\infty$~\cite{MorseFeshbach2}.
The Schr{\"o}dinger equation remains separable if we add a potential of the form
\begin{equation}
    V = \frac{(\xi_2^2-\xi_3^2)u(\xi_1)+(\xi_1^2-\xi_3^2)v(\xi_2)+(\xi_1^2-\xi_2^2)w(\xi_3)}{
      (\xi_1^2-\xi_2^2)(\xi_1^2-\xi_3^2)(\xi_2^2-\xi_3^2)}.
\end{equation}
The degenerate forms often lead to situations where two or more regular singular points
merge to irregular singular points (confluent singularities).

\subsection{The Fokker-Planck equation}

There are of course many other fields of physics where equations
similar to the Schr{\"o}dinger equation occur. One such example
is the Fokker-Planck equation~\cite{Fokker, MaxPlanck} describing
the time evolution of a probability distribution $\rho(\bm{r},t)$
of a diffusing particle in a force field $\bm{F}(\bm{r}) = -\bm{\nabla} U(\bm{r})$,
\begin{equation}
    \frac{\partial}{\partial \tau}\,\rho(\bm{r},\tau) = 
    \frac{1}{2}\,\bm{\nabla}^2\rho(\bm{r},\tau) - \bm{\nabla} \cdot \left[ \bm{F}(\bm{r}) \rho(\bm{r},\tau) \right].
\end{equation}
By writing $\rho(\bm{r}) = \text{e}^{-U(\bm{r})}\,\Psi(\bm{r},t)$ we obtain a 
``Schr{\"o}dinger equation'', in imaginary time $\tau = \text{i} t$, for $\Psi(\bm{r}, \tau)$,
\begin{equation}
   -\frac{\partial}{\partial\tau}\Psi(\bm{r},\tau) = 
   -\frac{1}{2} \,\bm{\nabla}^2\Psi(\bm{r},\tau)  +  V(\bm{r})\,\Psi(\bm{r},\tau),
\end{equation}
with a potential $V(\bm{r}) \equiv
\frac{1}{2}\left[\bm{F}(\bm{r})\!\cdot\!\bm{F}(\bm{r}) + \bm{\nabla}\!\cdot\!\bm{F}(\bm{r})\right]$.
For this reason we have denoted the equations we study
in Paper I \cite{HighPrecisionSolutions} and Paper II \cite{NormalizedEigenfunctions}
as \emph{Schr{\"o}dinger type} equations instead of just Schr{\"o}dinger equations.

\chapter{The projects of this thesis}

Although the projects of this thesis are inspired by all the points
in section~\ref{MostAccurate}, they are nearest to the first one. 
In the first project \cite{HighPrecisionSolutions}
the ground state energy of the anharmonic oscillator was computed
to the accuracy of one million decimal digits. There are certainly
no experimental results which require such accuracy, or physical
systems which is modeled to that accuracy by the given Hamiltonian.
That project was an attempt to explore the borders of applicability
of our method, using standard computers of the time.

The practical applications of very-high-precision calculations
is that they provide essentially exact results, and hence may
be useful for testing theoretical conjectures or the accuracy
of approximation methods. One may also think of practical problems
where numerical accuracy to several tens of decimals may
be useful.

\section{Properties of the used algorithms}

Most projects of this thesis are based on an algorithm for solving Schr{\"o}dinger
type equations to very high precision. 
Despite the numerical character of this topic, the focus of all investigations has been to
explore \emph{analytic} questions.
How do the required resources, like computer memory and CPU cycles,
scale with the wanted precision? With standard numerical methods, based on
discretization of the Schr{\"o}dinger differential operator, the error
$\varepsilon \equiv 10^{-D}$ scales like a low power of the discretization step \cite{AsifAnneKare},
\begin{equation*}
  \varepsilon \sim (\Delta x)^n,
\end{equation*}
usually with $1\le n \le ~6$. Hence the number $N$ of 
computational steps would grow {\em exponentially\/}
with the wanted precision,
\begin{equation*}
  N \sim 10^{D/n}.
\end{equation*}

In contrast, in the algorithm used here the number of
computational steps grows asymptotically {\em linearly\/} with the wanted precision $D$
as $D\to\infty$,
\begin{equation*}
   N \approx N_0 + \text{const}\times D.
\end{equation*}
This algorithm works for Schr{\"o}dinger type
equations with polynomial potentials in one dimension, and can be extended
to a much larger class of ordinary differential equations.
Note that in many situations the offset $N_0$ can be quite large,
making it computationally very expensive to obtain the
first few decimals of accuracy. Hence, our algorithm may not be suitable
for cases where only standard double-precision is wanted (but there are
examples where it is a competitive method). And we have not yet seriously explored the
possible improvements which may be made by appropriate use of analytic continuations.

\begin{figure}[H]
\begin{center}
\includegraphics[clip, trim= 10ex 5ex 10ex 1.75ex,width=0.90\textwidth]{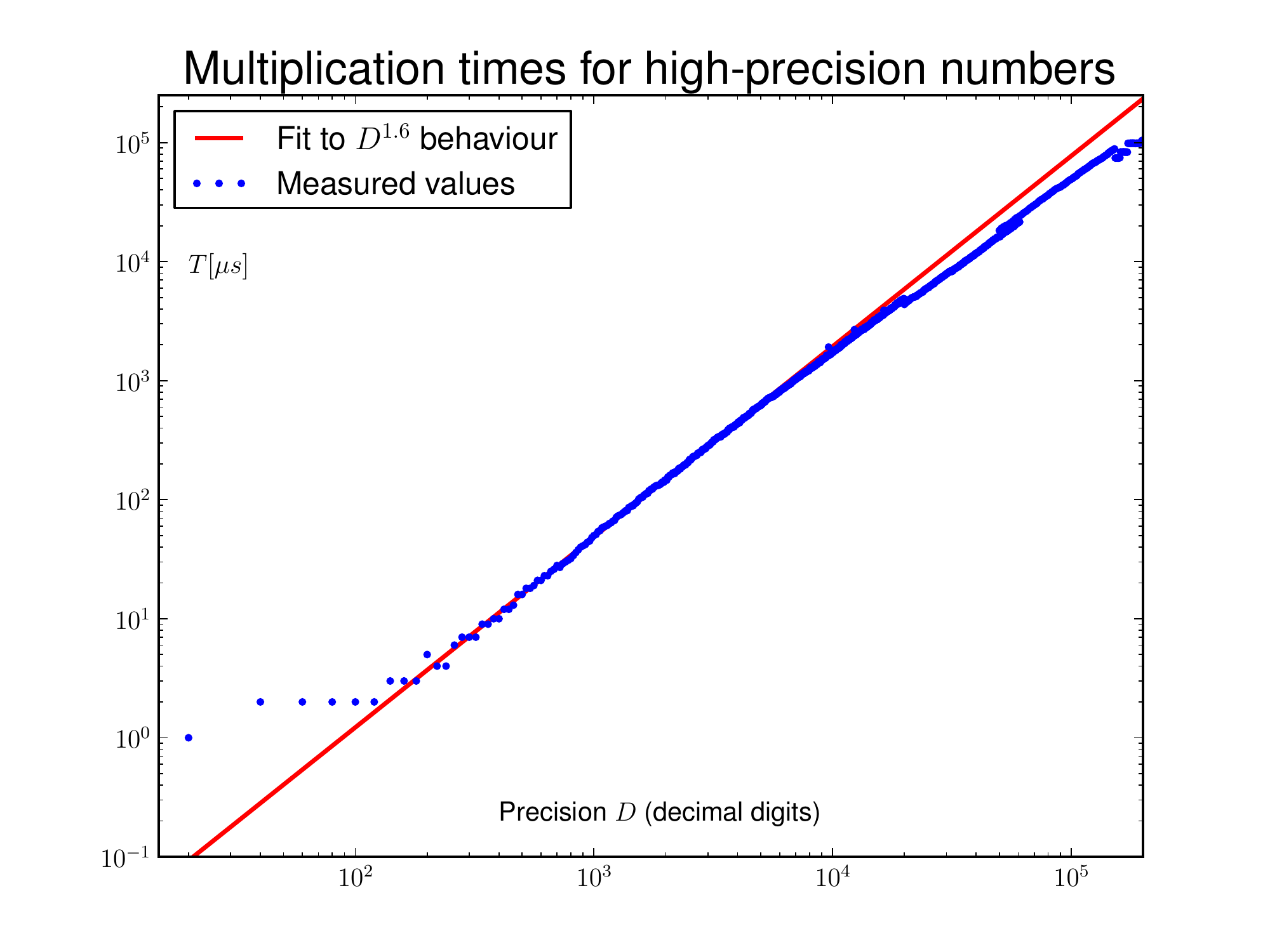}
\end{center}
\caption{The most time-consuming algebraic operation in our solution algorithm
is the multiplication of two high-precision numbers. This figure shows
a measurement of how this time varies with precision $D$ on a standard linux workstation.
The theoretical prediction is that it should grow like a power $D^{1.6}$ for low values
of $D$. As can be seen this is followed quite accurately for precision values in the range
$100 \le D \le 10^4$. Eventually the theoretical growth should reduce to a rate
$D\,\log D\,\log\log D$; this does not seem to be achievable in practice on our
computers. The discontinuous behavior at low $D$ occurs because the real 
computational precision increased in steps of $64$ bits, corresponding to $19^{+}$ decimal digits.
}\label{multiplication}
\end{figure}

It should be kept in mind
that each computational step will require more
time as one increases precision. In our method the multiplication of two high-precision
numbers is the most time-consuming operation.
We have based all
high-precision numerical computations on the CLN Class Library for Numbers \cite{CLN},
compiled with the GNU Multiple Precision Arithmetic Library \cite{GMP}.
These libraries
use the Sch{\"o}nhage-Strassen algorithm \cite{SchonhageStrassen} for
multiplying numbers. With this algorithm the time $T$ of multiplying two
high precision numbers scale theoretically at a rate between $D^{1.6}$ and $D\,\log D\,\log\log D$
with precision $D$. Hence the total time requirement of our high-precision method
increases somewhat faster than $D^2$ with the precision $D$. In practical computations
one finds $T \sim D^{1.6}$ for $D \le 2000$, cf.~figure~\ref{multiplication}.

Although it is of limited practical interest to compute physical quantities
like eigen-energies to much more than 20 decimals precision, it may be the
case that computation of the wave-functions in some regions of interest require
very precise knowledge of the eigen-energy --- otherwise it would be impossible to
find solutions with the correct behavior. This is one practical motivation for
developing methods to solve eigenvalue problems to otherwise
ridiculously high precision.

In paper II \cite{NormalizedEigenfunctions} it was demonstrated,
and analyzed, that the very-high-precison wave-functions can be inserted into
straightforward numerical integration routines to compute normalization
integrals to comparable precision within acceptable time.

We believe that these algorithms also can be used to compute
functional determinants of one-dimensional differential
operators to comparable precision. Since the functional determinant of a
separated sum of differential operators is the product of the individual determinants, 
the algorithms can also be used to compute functional
determinants of higher-dimensional separable differential operators \cite{Jentschura_ZinnJustinI}.
The algorithms can also be used to compute the resolvent of one-dimensional
differential operators, but in this case an extension to
higher-dimensional separable operators becomes more complicated since one
must first compute the partition functions of the individual one-dimensional operators.

\section{Method for solving the differential equation\label{Method}}

Our method of solution is (perhaps disappointingly) simple:
We just expand the solution $\psi(x)$ in a power series,
\begin{equation}
    \psi(x) = x^{\sigma} \sum_{m\ge 0} a_m x^m,
    \label{PowerExpansion}
\end{equation}
where $\sigma=0$ for the even parity solutions, and $\sigma=1$ for the odd parity solutions.
For a given $x$ the quantities $A_m(x) \equiv a_m x^{2m+\sigma}$ can be generated recursively from
equation~(\ref{1dSchrodingerEquation}),
\begin{equation}
   A_{m+1}(x) =  \frac{1}{(2m+2+\sigma)(2m+1+\sigma)} \sum_{n=0}^N V_n(x)\,A_{m-n}(x),
   \label{RecursionRelation}
\end{equation}
where 
\begin{equation}
  V_n(x)= \left\{
    \begin{array}{ll}
    v_0-\varepsilon, &\text{for } n=0,\\
    v_n\, x^{2n},       &\text{for } 1\le n < N,\\
    x^{2N},           &\text{for } n=N.
      \end{array}
  \right.
\end{equation}
This recursion is initialized with $A_{-n}(x) = \delta_{n0}\, x^{\sigma}$ ($n\ge0$). 
As is seen from equation~(\ref{RecursionRelation}) only the last $N+1$ coefficients $A_m(x)$
need to be considered at any time while the sum is accumulated. 
This means that one only needs to store
$N+1$ coefficients $A_{m}$ and at most $N+1$ coefficients $V_{n}$, together with the accumulated sums
for $\psi(x)$ and optionally $\psi'(x)$. Hence the total memory requirement is $2N+4$ high precision numbers,
which is quite modest when the potential is a low order polynomial.

In principle the sum in equation (\ref{PowerExpansion}) runs over infinitely many terms. But
since equation~(\ref{1dSchrodingerEquation}) has no singular points in the finite plane the sum will
eventually converge very fast, and can be cut off at some finite value ${\cal N}$. 
The value of ${\cal N}$ depends on the desired accuracy
of the sum, the value of $x$, and of course the parameters in equation~(\ref{PowerExpansion}).

The energy eigen-function~$\psi(x)$ should become very small when $\vert x \vert$ becomes very large.
Hence there will be large cancellations in the sum, and associated accumulation of
round-off errors. Therefore, the interesting aspects of this method is not to code the recursion
algorithm~(\ref{RecursionRelation}), but f.i.~to estimate the proper choice of ${\cal N}$ and
the numerical precision which must be employed in the computation.
We have found that the WKB approximation combined with a Legendre transformation
can be used for this analysis, see article IV(a) \cite{WKBLegendre}.

\section{Eigenvalue conditions}

The standard eigenvalue condition is that
$\psi(x)\to 0$ as $x\to\pm\infty$. With parity symmetry ($x\to -x$) it is
sufficient to consider only one of these limits. I.e., we may start with solutions
$\psi_{\sigma}(x; \varepsilon)$ of desired parity, and use the single condition
\begin{equation}
    \lim_{x\to\infty} \psi_{\sigma}(x,\varepsilon) = 0,
    \label{SymmetricCondition}
\end{equation}
to determine the allowed eigenvalues $\varepsilon$.

The assumption that the potential $V$ should be even is not necessarily physical,
and not really necessary. In the general case one can expand the
general solution in two linearly independent solutions $\varphi_a(x; \varepsilon)$ and
$\varphi_b(x; \varepsilon)$,
\begin{equation*}
   \psi(x; \varepsilon) = C_a\,\varphi_a(x; \varepsilon) + C_b\,\varphi_b(x; \varepsilon),
\end{equation*}
where $C_a$ and $C_b$ are coefficients to be determined. The eigenvalue condition is then
that there should be a nontrivial solution of the equation
\begin{equation*}
    \lim_{x\to\infty}\left(\begin{array}{cc}\varphi_a(x;\varepsilon)&\varphi_b(x;\varepsilon)\\
        \varphi_a(-x;\varepsilon)&\varphi_b(-x;\varepsilon)\end{array}\right) 
    \left(\begin{array}{c} C_a\\C_b\end{array}\right)  = 0.
\end{equation*}
The condition for this is that the system determinant must vanish,
\begin{align}
   &\lim_{x\to\infty}\text{det}\left(\begin{array}{cc}\varphi_a(x;\varepsilon)&\varphi_b(x;\varepsilon)\\
       \varphi_a(-x;\varepsilon)&\varphi_b(-x;\varepsilon)\end{array}\right)
   =\nonumber\\[-2.5ex]
   \label{AsymmetricCondition}\\
   &\lim_{x\to\infty} \left[\,\varphi_a(x;\varepsilon)\,\varphi_b(-x;\varepsilon) - 
     \varphi_a(-x;\varepsilon)\, \varphi_b(x; \varepsilon)\,\right] = 0.
   \nonumber
\end{align}

This is in principle not different
from the condition that a single function must vanish,
but is in practice a nontrivial extension. The numerical
implementation becomes more complex, since we must solve
an eigenvalue problem, and the program execution will be more demanding
since one must compute and store four functions at each step
instead of one. The assumption that the polynomial $V$ should
be of low order, i.e.~that $N$ should be small, is mostly
motivated by storage requirement.

With our method of solution it is not possible to evaluate
the wave-functions at, or very close to, $x=\infty$. We
must therefore replace the condition~(\ref{SymmetricCondition})
or (\ref{AsymmetricCondition}) with boundary conditions at
finite, but sufficiently large, $x_0$. For each eigenvalue
$\varepsilon$ there exist a Robin boundary condition at $x_0$,
\begin{equation}
    -\frac{\psi'(x_0)}{\psi(x_0)} = R(x_0)
\end{equation}
which is equivalent to (\ref{SymmetricCondition}). We do not
know $R(x_0)$ exactly (it even depends weakly on
the eigenvalue $\varepsilon$), but it can be estimated from asymptotic
analysis of equation~(\ref{1dSchrodingerEquation})
as $x_0\to\infty$. Define
\begin{equation*}
    Q(x) \equiv x^{2N} + \sum_{n=0}^{N-1}\,v_n\,x^{2n} - \varepsilon,
\end{equation*}
and let $\bar{x}$ be the largest solution of $Q(x)=0$ (i.e., the largest
classical turning point). Since $Q(x)$ becomes large when $x\gg \bar{x}$
we may choose
\begin{equation}
     R(x_0) = \sqrt{Q(x_0)} + \text{ higher-order corrections}
     \label{RobinParameter}
\end{equation}
for a sufficiently large $x_0 \gg \bar{x}$. This may often be further approximated by
$R(x_0) = \infty$. I.e., a Dirichlet boundary condition, $\psi(x_0)=0$, at $x_0$.
The most important property is that the eigenvalue is not very sensitive to the
precise choice of $R(x_0)$. For $x > \bar{x}$ the solution can be approximated as
\begin{align}
   \psi(x) \approx\;  &K_{+}(\varepsilon)\,Q(x)^{-1/4}\,\exp\left({\int_{\bar{x}}^x dt \sqrt{Q(t})}\right) +\nonumber\\[-2ex]
   \\
   &K_{-}(\varepsilon)\,Q(x)^{-1/4}\,\exp\left({-\int_{\bar{x}}^x dt \sqrt{Q(t})}\right),\nonumber
\end{align}
where the coefficients $K_{\pm}$ are expected to be of the same magnitude in general
(both of order unity and slowly varying with $\varepsilon$).

The exact quantization condition, the Dirichlet boundary condition at $x_0$, and the Robin boundary
condition at $x_0$ become respectively, with $Q_0 \equiv Q(x_0)$ and $Q'_0 \equiv Q'(x_0)$, 
\begin{subequations}
  \label{BoundaryConditions}
  \begin{alignat}{1}
     &K_{+}(\varepsilon) = 0,\label{Exact}\\
     &K_{+}(\varepsilon) + K_{-}(\varepsilon)\,
     \exp\left({-2\int_{\bar{x}}^{x_0} dt \sqrt{Q(t})}\right)= 0,\label{Dirichlet}\\
     &K_{+}(\varepsilon) - \frac{1}{8}K_{-}(\varepsilon)\,Q_0^{-3/2} Q'_0\,
     \exp\left({-2\int_{\bar{x}}^{x_0} dt \sqrt{Q(t})}\right)= 0.\label{Robin}
   \end{alignat}
\end{subequations}
To compare solutions let $\varepsilon_{\text{exact}}$ be the solution of equation~(\ref{Exact}), and expand
\begin{equation*}
    K_+(\varepsilon) = K'_+(\varepsilon_{\text{exact}})\,\left(\varepsilon-\varepsilon_{\text{exact}}\right) + \ldots.
\end{equation*}
Then the solutions of equations~(\ref{Dirichlet}) and (\ref{Robin}) become respectively
\begin{subequations}
  \begin{alignat}{1}
    \varepsilon_{\text{Dirichlet}} &= \varepsilon_{\text{exact}} - \frac{K_{-}(\varepsilon_{\text{exact}})}{K'_+(\varepsilon_{\text{exact}})}\,
    \exp\left({-2\int_{\bar{x}}^{x_0} dt \sqrt{Q(t})}\right),\\
    \varepsilon_{\text{Robin}} &= \varepsilon_{\text{exact}} + \frac{K_{-}(\varepsilon_{\text{exact}})}{8\,K'_+(\varepsilon_{\text{exact}})}
    \,Q_0^{-3/2} Q'_0\,\exp\left({-2\int_{\bar{x}}^{x_0} dt \sqrt{Q(t})}\right).
    \end{alignat}
\end{subequations}
%\begin{wrapfigure}[13]{r}{0.5\linewidth}
%\end{wrapfigure}

\begin{figure}[H]
\begin{center}
\includegraphics[clip, trim= 10ex 5ex 10ex 1.75ex,width=0.90\textwidth]{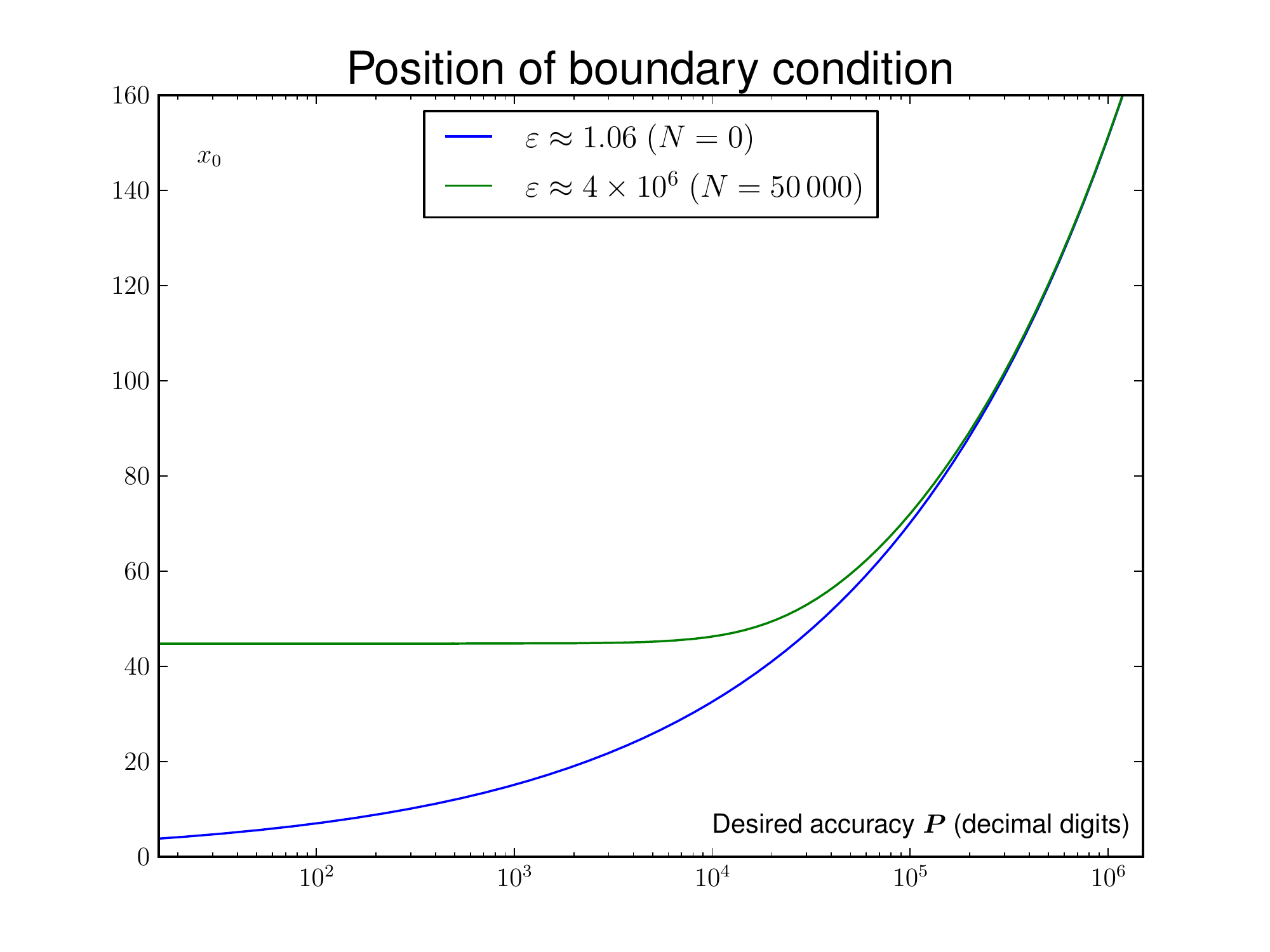}
\end{center}
\caption{The exact boundary condition, that $\vert \psi(x) \vert \to 0$ as $\vert x \vert \to \infty$,
cannot be realized in a numerical computation, but must be replaced by a
Robin or Dirichlet boundary condition at some finite (large) value $x_0$.
This figure illustrates how $x_0$ must be chosen to maintain a described accuracy $10^{-P}$ of the
computed eigenvalues, for the pure anharmonic oscillator, $V(x) = x^4$, for the lowest eigenvalue
and eigenvalue number $N=50\,000$. Large values of $x_0$ is computationally challenging, but possible,
even if one evaluates the wave-function directly by a series expansion around $x=0$.\label{PositionOfBoundary}
}
\end{figure}
We note that the Robin boundary condition improves the
accuracy by a relative amount $\frac{1}{8}Q_0^{-3/2} Q'_0$,
which may correspond to a few decimal digits. This may be
further improved by systematically adding correction terms
to the parameter $R(x_0)$, cf.~equation~(\ref{RobinParameter}).
Each order of correction will provide a few extra decimals
of accuracy. This may be a good approach if one wants the
eigenvalue to standard precision (i.e.~15-16 decimal digits)
only, but it makes little difference
if one wants hundreds of
decimals or more. Then one must make use of the fact that
the factor\footnote{For this crude estimate of the integral we
make a partial integration, using $Q(\bar{x})=0$, and the approximation $t\frac{d}{dt} Q(t) \approx 2N\,Q(t)$.}
\begin{equation}
  \exp\left({-2\int_{\bar{x}}^{x_0} dt \sqrt{Q(t})}\right)
  \sim \exp\left(-\frac{2x_0\, Q^{1/2}_0 }{N+1} \right)
\end{equation}
vanishes exponentially fast
with increasing $x_0$. As an example consider the case that $Q(x) = x^4-\varepsilon$.
Then the quantization condition (\ref{Dirichlet}) or (\ref{Robin}) leads
to an error in the eigenvalue of order
\begin{equation}
    \Delta\varepsilon \sim \text{e}^{-\frac{2}{3} x_0 \sqrt{x_0^4-\varepsilon}},
\end{equation}
which clearly vanishes exponentially fast as $x_0$ increases.
Figure~\ref{PositionOfBoundary} displays how large one must
select $x_0$ to obtain a desired accuracy of
eigenvalue number $N$. As can be seen, this value is quite
large for large $N$, even at moderate accuracy.
This may not be a serious obstacle if $\psi(x)$
is evaluated by analytic continuation, but it is
quite time-consuming if one evaluates it by a direct
power series expansion around the origin.

\part{Mathematical background}

\chapter{Linear Ordinary Differential Equations}

In this part the mathematical background used for the thesis projects
is discussed. Some of the more technical parts is placed in the
Appendices.

A linear differential equation is any differential equation that can be written 
in the following form,
\begin{equation}
  \left [p_{n}(z)\frac{d^{n}}{dz^{n}} + p_{n-1}(z)\frac{d^{n-1}}{dz^{n-1}} + \cdots + p_{0}(z)\right ]f(z) = g(z),
  \label{Linearequation}
\end{equation}
or more symbolically,
\begin{equation}
  {\cal L}f = g.
\end{equation}
Equation~(\ref{Linearequation}) is said to be of order $n$,
it is called linear because ${\cal L}$ is a linear operator,
\begin{equation}
  {\cal L}(c_{1}f_{1} + c_{2}f_{2}) = c_{1}{\cal L}f_{1} + c_{2}{\cal L}f_{2}
\end{equation}
when $c_{1}$ and $c_{2}$ do not depend on $z$,
and ordinary because it only involves one independent
variable as opposed to a partial differential equation.
If $g=0$ the equation is said to be homogeneous, 
otherwise it is inhomogeneous.

\section{Expansions around ordinary and regular singular points}

We will assume the coefficients $p_k(z)$ to be analytic (usually polynomials)
in the region of interest.
The homogeneous version of equation~(\ref{Linearequation})
is said to have a {\em regular singular point\/}
at $z_0$ if some of the functions $p_k(z)/p_n(z)$ has a (perhaps higher
order) pole singularity at $z_0$, but such that each
\begin{equation}
   \alpha_k \equiv \lim_{z\to z_0}  (z-z_0)^{n-k} p_k(z)/p_n(z)
\end{equation}
is finite. In generic cases the solutions can be expanded in Frobenius series
around $z_0$,
\begin{equation}
    f(z) = (z-z_0)^{\sigma} \sum_{m\ge 0} f_m\,(z-z_0)^m,
    \label{Frobenius_Series}
\end{equation}
provided $\sigma$ is a solution of the {\em indicial equation\/}
\begin{equation}
    \sum_{k=0}^n \alpha_k\, \sigma^k = 0.
    \label{indicial_equation}
\end{equation}
However, there are exceptional cases when equation~(\ref{indicial_equation}) has multiple
roots, or when some roots differ by integers, for which one must modify
the series (\ref{Frobenius_Series}) with logarithmic factors.
The general analysis becomes rather complicated, with many special cases to consider.
For this reason the published code in Paper III \cite{SeriesSolutions}
excludes all exceptional cases.

For second order equations the exceptional cases can be reduced to only
two possibilities. Since it was difficult to find 
the general recursion formulas for even these cases in
the literature, in an explicit form suitable for coding,
we have derived and published them in Paper IV(b) 
\cite{GeneratingFrobeniusSeries}.

\section{Second-order equations as first-order systems}

Many ordinary differential equations originate  by
separation of variables from partial differential equations
involving the Laplace operator, and will therefore be of second
order. They can often be transformed to the form
\begin{equation}
   \left[p(z)\,\frac{d^2}{dz^2} + q(z)\,\frac{d}{dz} + r(z)\right] f(z) = 0,
   \label{SecondOrderODE}
\end{equation}
where $p(z)$, $q(z)$, and $r(z)$ are low-order polynomials in $z$,
or the inhomogeneous version of such equations. This equation
is well suited for expansion in a Frobenius series
(cf.~equation~(\ref{PowerExpansion})), 
\begin{equation*}
    \psi(z) = z^{\sigma} \sum_{m\ge 0} a_m z^m,
\end{equation*}
since only a few coefficients $a_{m-k}$ are
required for computation of each next coefficient $a_{m+1}$. Although the series
is assured to converge up to the nearest singular point of equation~(\ref{SecondOrderODE}), 
it may be convenient to evaluate $f(z)$ indirectly through one or more points $z_i$ by
analytic continuation. Analytic continuation of functions which satisfy a
second-order differential equation is rather simple to implement, since the function
is fully specified by just two complex numbers, $f(z_i)$ and $f'(z_i)$, together with the
differential equation.

Hence we want to make coordinate transformations of
equations like (\ref{SecondOrderODE}), and to implement robust
algorithms for such transformations. This is simple as
long as we limit ourselves to translations, since a
translation of the independent variable, $z = z_0 + u$, only
transforms equation (\ref{SecondOrderODE}) to
\begin{equation*}
   \left[\tilde{p}(u)\,\frac{d^2}{du^2} + \tilde{q}(u)\,\frac{d}{du} + \tilde{r}(u)\right] \tilde{f}(u) = 0,
\end{equation*}
where $\tilde{p}(u) = p(z_0+u)$, $\tilde{q}(u)= q(z_0+u)$, $\tilde{r}(u)= r(z_0+u)$ are polynomials of
the same order as the original ones. However, we would like to include the full group of
M{\"o}bius transformations,
\begin{equation}
    z = \frac{\alpha u + \beta}{\gamma u + \delta},
    \quad\text{with } \alpha \delta - \beta \gamma = 1.
    \label{MobiusTransformation}
\end{equation}
These are the most general transformations of the Riemann sphere
which do not introduce new singularities. It is not straightforward to describe
a class of equations of the form~(\ref{SecondOrderODE}) which are invariant under
M{\"o}bius transformations, in particular to computers. For this reason it seems preferable
to reformulate (\ref{SecondOrderODE}) as a system of first-order equations. This should be
done with some care. The perhaps most obvious choice
\begin{equation*}
    y_0(z) = f(z),\quad
    y_1(z) = f'(z),
\end{equation*}
introduces an irregular singular point $z=\infty$, since (with $u=1/z$)
\begin{equation*}
    \frac{d}{du}y_0(u^{-1}) = -\frac{1}{u^2} y_1(u^{-1}).
\end{equation*}
This is unwanted unless $z=\infty$ already is an irregular singular point.
Instead, if equation~(\ref{SecondOrderODE}) has singular points at $z=z_0$ and $z=z_1 \ne \infty$,
the choice
\begin{equation*}
    y_0(z) = f(z),\quad y_1(z) = C\,(z-z_0)(z-z_1)f'(z),    
\end{equation*}
will not introduce new singularities. One may choose the constant $C\ne 0$ freely.
By setting $C=-z^{-1}_1$ and taking the limit $z_1\to\infty$ one obtains
\begin{equation*}
    y_0(z) = f(z),\quad y_1(z) = (z-z_0)f'(z).
\end{equation*}

\subsection{Example: Reformulation of the hypergeometric equation}

Consider the hypergeometric equation
\begin{equation}
    z(1-z)f''(z) + \left[c-(a+b+1)z \right] f'(z) - ab\, f(z) = 0.
\end{equation}
This is known to have regular singular points at $z=0, 1,\infty$,
and no other singular points. Hence, there are three possible
pairs of singular points which may be used to define $y_1(z)$.
We choose $z=0$ and $\infty$, and a vector $\bm{y}$ with components
\begin{equation}
    y_0 = f, \quad y_1 = z f',
\end{equation}
and obtain the system of first-order equations
\begin{equation}
    z(1-z)\frac{d}{dz}
    \begin{pmatrix}
      y_0\\ y_1
    \end{pmatrix}
    =
    \begin{pmatrix}
      0&1-z\\
      ab\,z&1-c +(a+b)z
    \end{pmatrix}
    \begin{pmatrix}
      y_0\\y_1
    \end{pmatrix}.
\end{equation}
The regular singular points of this equation, with the corresponding indices,
can be arranged according to the \emph{tableau}
\begin{equation*}
  \begin{Bmatrix}
     0&1&\infty\\
     0&0&a\\
     1-c&c-a-b-1&b
  \end{Bmatrix}.
\end{equation*}
The first line of this pattern lists the positions of the regular singularities
(here $z=0$, $z=1$, and $z=\infty$), and the two entries below each position
are the indices $\nu_{k}$ at that position. 
Here $\nu_1=0$, $\nu_2 = 1-c$ at $z=0$, 
$\nu_1 = 0$, $\nu_2 = c-a-b-1$ at $z=1$, and
$\nu_1 = a$, $\nu_2 = b$ at $z=\infty$.
Note that the sum of all indices at all
singular points is zero, not equal to one as in the 2nd order formulation.

The general equation
with regular singular points at $0, 1, \infty$ is obtained by
considering $\bm{f}(z) = z^\mu (1-z)^\nu \bm{y}(z)$. Since
\begin{equation*}
     z(1-z) \frac{d}{dz} z^{\mu} (1-z)^{\nu} = -z^{\mu} (1-z)^{\nu}\left[\mu(z-1)+\nu z \right],
\end{equation*}
one finds
\begin{equation}
    z(1-z)\frac{d}{dz} \bm{f}(z) =
    \begin{pmatrix}
      \mu\!-\!(\mu\!+\!\nu)z&1\!-\!z\\
      ab\,z&(1\!+\!\mu\!-\!c)\!+\!(a\!+\!b\!-\!\mu\!-\!\nu)z
    \end{pmatrix}
    \bm{f}(z).
\end{equation}
The right-hand side is still a first-order matrix polynomial,
in contrast to the second-order formulation where the polynomials increase in order
when one generalizes the hypergeometric equation in the same manner.

Finally, a M{\"o}bius transformation~(\ref{MobiusTransformation})
which transforms the points
$0, 1, \infty$ to $u_0, u_1, u_\infty$ leads to the equation

{\footnotesize
\begin{align}
    &R_1 (u-u_0)(u-u_1)(u-u_{\infty}) \frac{d}{du} \bm{\tilde{f}}(u)
    \nonumber\\&=
    \begin{pmatrix}
      -\nu R_2(u\!-\!u_0)+\mu (u_1\!-\!u)&u_1\!-\!u\\
      ab R_2(u\!-\!u_0)&\!\!\!\!\!\!\!\!(1\!+\!\mu\!-\!c) R_3(u\!-\!u_{\infty}) \!+\! (a\!+\!b\!-\!\mu\!-\!\nu)R_2(u\!-\!u_0)
    \end{pmatrix}
    \bm{\tilde{f}}(u),
\end{align}
}
\noindent
with
\begin{equation}
    R_1 = \left(\frac{1}{u_\infty-u_0}\right),\quad
    R_2 = \left(\frac{u_1-u_{\infty}}{u_0-u_{\infty}}\right),\quad
    R_3 = \left(\frac{u_1-u_0}{u_0-u_\infty}\right).
\end{equation}
The important feature here is that the factor multiplying $\frac{d}{du}$ is a third-order polynomial in $u$,
and that right-hand side involves a first-order matrix polynomial in $u$. This structure is stable under
M{\"o}bius transformations, and transformations
\begin{equation*}
 \bm{f}(u) \to (u-u_0)^{\mu}(u-u_1)^{\nu}(u-u_{\infty})^{-\mu-\nu} \bm{f}(u).
\end{equation*}

\section{System of first-order equations}

Consider a system of first-order equations,
\begin{equation}
    p(z) \frac{d}{dz}\bm{y}(z) = \bm{A}(z)\,\bm{y}(z),
    \label{FirstOrderEquation}
\end{equation}
where $\bm{y}(z)$ is a $K$-component vector, $p(z)$ is an ordinary polynomial
of order $N$, and $\bm{A}(z)$ is a $K\times K$ matrix polynomial,
\begin{align}
    p(z) &= \sum^N_{k=0} p_k\,z^k \equiv C\,\prod_{k=0}^{N-1} (z-z_k),\\
    \bm{A}(z) &= \sum_{k\ge0} \bm{A}_k\,z^k.
\end{align}

\subsection{Expansion around an ordinary point}

If $p_0 \ne 0$ the solution can be expanded in an ordinary power series,
\begin{equation}
    \bm{y}(z) = \sum_{m\ge 0}  \bm{a}_m\,z^m.
    \label{TaylorSeries}
\end{equation}
We insert the series into equation~(\ref{FirstOrderEquation}) and introduce matrices
\begin{equation}
   {\cal M}_k(\mu) \equiv (\mu - k)\,p_{k+1} - \bm{A}_k.
\end{equation}
In terms of these one finds the recursion formula
\begin{equation}
     \bm{a}_{m+1} = -\frac{1}{(m+1) p_0} \sum_{k\ge 0} {\cal M}_k(m)\, \bm{a}_{m-k},
\end{equation}
where $\bm{a}_0$ can be chosen freely, and $\bm{a}_{-n} = 0$ for $n=1, 2,\cdots$. According to general theory
the series~(\ref{TaylorSeries}) will converge at least to the closest zero of $p(z)$. I.e., the
radius $R$ of convergence satisfies
\begin{equation}
      R \ge \min_k \,\vert z_k \vert. 
\end{equation}

\subsection{Expansion around a regular singular point}

If $p_0=0$ but $p_1\ne 0$ the point $z=0$ is a regular singular
point for equation~(\ref{FirstOrderEquation}). The solution can be
found by use of the Frobenius method. We assume a solution of the form
\begin{equation}
    \bm{y}(z; \nu) = \sum_{m\ge 0}  \bm{a}_m\,z^{m+\nu},
    \label{FrobeniusSeries}    
\end{equation}
and find that equation~(\ref{FirstOrderEquation}) implies
\begin{equation}
     \left(p(z)\frac{d}{dz} -\bm{A}\right) \bm{y}(z; \nu) = 
     \sum_{m\ge 0} \left( \sum_{k\ge 0} {\cal M}_k(m+\nu)\,\bm{a}_{m-k}\right) z^{m+\nu} = 0,
\end{equation}
where the coefficients $\bm{a}_{-n} =0$ for $n=1,2,\cdots$.
The coefficient of each power $z^{m+\nu}$ must vanish. For $m=0$
this implies
\begin{equation}
    {\cal M}_0(\nu) \bm{a}_0 = 0,
\end{equation}
which has a nontrivial solution only when the indicial equation,
\begin{equation}
    \text{det}{\cal M}_0(\nu)  = 0,
    \label{IndicialEquation}
\end{equation}
is fulfilled. This $K^{\text{th}}$-order algebraic equation has $K$
solutions $\nu_s$ (counting multiplicities). There is at least
one right eigenvector $\bm{a}_0(\nu_s)$ for each distinct index $\nu_s$,
and equally many left eigenvectors $\bm{\bar{a}}_0(\nu_s)$.
Each index $\nu_s$ corresponds to an eigenvalue
$\lambda_s = \nu_s p_1$ of the matrix $\bm{A}_0$.

\subsection{Distinct indices with non-integer differences}

Assume first that all indices $\nu_s$ are distinct, and that
the difference between any two of them is non-integer. Then higher-order
(vector) coefficients can be computed recursively as
\begin{equation}
     \bm{a}_{m+1}(\nu) = -{\cal M}_{0}(m+1+\nu)^{-1}\,\sum_{k\ge1} {\cal M}_k(m+1+\nu)\,\bm{a}_{m+1-k}(\nu),
\end{equation} 
for $m=1,2,\cdots$,
where the coefficients $\bm{a}_{-n} =0$ for $n=1,2,\cdots$.
The recursion is solvable at each step since all matrices ${\cal M}_{0}(m+1+\nu)$ are invertible by assumption.
(If one of them were not, the corresponding $m+1+\nu$ would also be a solution of (\ref{IndicialEquation}), contrary
to the assumption.)

\subsection{Distinct indices; one pair with integer difference}

Assume next that all indices $\nu_s$ are distinct, that
the difference between two of them is integer, 
$\nu_2 = \nu_1 + \ell$ with $\ell>0$,
and that all other possible differences are non-integer.
We must then make a more general solution ansatz for the
$\nu_1$-solution,
\begin{equation}
     \bm{y}(z;\nu_1) = \sum_{m\ge 0} \bm{a}_{m}\,z^{m+\nu_1} + \sum_{m\ge \ell} \bm{b}_{m}\,z^{m+\nu_1}\,\log(z).
\end{equation}
Now equation~(\ref{FirstOrderEquation}) implies
\begin{align}
     &\left(p(z)\frac{d}{dz} -\bm{A}\right) \bm{y}(z; \nu_1) = 
     \sum_{m\ge 0} \sum_{k\ge 0} {\cal M}_k(m+\nu_1)\,\bm{a}_{m-k}\, z^{m+\nu_1} \nonumber\\[-2.5ex]
     \\
     &+ \sum_{m\ge \ell} \sum_{k\ge 0} p_{k+1}\bm{b}_{m-k}\, z^{m+\nu_1} + {\cal M}_k(m+\nu_1)\,\bm{b}_{m-k}\, z^{m+\nu_1}\,\log(z) =0.
     \nonumber
\end{align}
The coefficients of each of the terms $z^{m+\nu}$ and $z^{m+\nu}\,\log(z)$ must vanish.
For $m=0$ this implies
\begin{equation}
    {\cal M}_0(\nu_1) \bm{a}_0 = 0,   
\end{equation}
as before. For $m=\ell$ this implies
\begin{align}
    {\cal M}_0(\ell + \nu_1) \bm{b}_{\ell} = 0,   
\end{align}
which has a nontrivial solution since $\ell + \nu_1 = \nu_2$ is
also an index by assumption. Hence $\bm{b}_{\ell}$ must be chosen
as a right eigenvector corresponding to the index $\nu_2$. There
is a corresponding left eigenvector $\bm{\bar{b}}_\ell$. For $m=\ell$
we must further have
\begin{equation}
   {\cal M}_0(\ell+\nu_1)\,\bm{a}_{\ell} =  -p_1\,\bm{b}_{\ell} -\sum_{k\ge 1} {\cal M}_k(\ell+\nu_1)\,\bm{b}_{\ell-k}.
   \label{EllEquation}
\end{equation}
This is not solvable for $\bm{a}_{\ell}$ in general, since $\text{det}\,{\cal M}_0(\ell+\nu_1)=0$.
The solution criterion is that the right-hand side must be
orthogonal to the left eigenvector $\bm{\bar{b}}_\ell$, i.e. that
\begin{equation}
   p_1\,\bm{\bar{b}}_\ell\cdot \bm{{b}}_\ell + \sum_{k\ge 1} \bm{\bar{b}}_\ell\cdot{\cal M}_k(\ell+\nu_1)\,\bm{b}_{\ell-k} = 0.
\end{equation}
We can always choose the length of $\bm{{b}}_\ell$ such that this equation is fulfilled. Then 
(\ref{EllEquation}) can be solved for $\bm{a}_{\ell}$. The solution is not unique. We can
add a right eigenvector $\bm{{b}}_\ell$ of arbitrary length to $\bm{a}_{\ell}$, and still
have a solution. This corresponds to adding a solution proportional to
\begin{equation*}
    \bm{y}(z,\nu_2) = \sum_{m\ge 0} \bm{a}_m(\nu_2)\,z^{m+\nu_2}.
\end{equation*}
The remaining coefficients can then calculated by the recursion formulas
\begin{equation}
    \bm{a}_{m+1}(\nu_1) = -{\cal M}_{0}(m+1+\nu_1)^{-1}\,\sum_{k\ge1} {\cal M}_k(m+1+\nu_1)\,\bm{a}_{m+1-k}(\nu_1),
\end{equation}
for $m+1 = 1, 2,\cdots,\ell-1$, and
{\footnotesize
\begin{align}
    \bm{b}_{m+1}(\nu_1) &= -{\cal M}_{0}(m+1+\nu_1)^{-1}\,\sum_{k\ge1} {\cal M}_k(m+1+\nu_1)\,\bm{b}_{m+1-k}(\nu_1),\nonumber\\[-2.5ex]
    &\\
    \bm{a}_{m+1}(\nu_1) &= -{\cal M}_{0}(m+1+\nu_1)^{-1}\,\sum_{k\ge1} {\cal M}_k(m+1+\nu_1)\,\bm{a}_{m+1-k}(\nu_1)\nonumber 
    + p_{k+1}\,\bm{b}_{m+1-k}(\nu_1),
\end{align}
}
\noindent
for $m+1 = \ell+1, \ell+2, \cdots$.

\subsection{One double degenerate index}

Assume next that an index $\nu_1$ is doubly degenerate,
that all the others are distinct, and that all index differences
are non-integer. If there are two linearly independent right eigenvectors $\bm{a}_{m}(\nu_1,r)$
corresponding to the index $\nu_1$, then the solution ansatz (\ref{FrobeniusSeries})
still works:
\begin{equation}
     \bm{y}(z;\nu_1) = \sum_{m\ge 0} \sum_{r=1}^2 \bm{a}_{m}(\nu_1,r)\,z^{m+\nu_1}.
\end{equation}
If there is only one eigenvector one again makes an ansatz with a logarithmic term
\begin{equation}
     \bm{y}(z;\nu_1) = \sum_{m\ge 0} \bm{a}_{m}(\nu_1)\,z^{m+\nu_1} + \bm{b}_{m}(\nu_1)\,z^{m+\nu_1}\,\log(z).
\end{equation}
Equation~(\ref{FirstOrderEquation}) implies that
\begin{align}
    &{\cal M}_0(m+\nu_1)\,\bm{b}_m + \sum_{k\ge 1} {\cal M}_k(m+\nu_1)\,\bm{b}_{m-k} = 0,\nonumber\\[-2.5ex]
    \\
    &{\cal M}_0(m+\nu_1)\,\bm{a}_m + p_1\,\bm{b}_m + 
    \sum_{k\ge 1} {\cal M}_k(m+\nu_1)\,\bm{a}_{m-k} + p_{k+1}\,\bm{b}_{m-k}= 0,\nonumber
\end{align}
for $m=0, 1, \cdots$, with $\bm{b}_{-n} = \bm{a}_{-n} = 0$ for $n=1, 2, \cdots$.
 For $m=0$ this implies that
\begin{equation}
    {\cal M}_0(\nu_1) \,\bm{b}_0 = 0,
\end{equation}
i.e. that $\bm{b}_0$ must be a right eigenvector corresponding to the index $\nu_1$, and
\begin{equation}
    {\cal M}_0(\nu_1) \,\bm{a}_0 = -p_1\,b_{0}.
\end{equation}
Even though $\text{det}\, {\cal M}_0(\nu_1)=0$, this equation does have a
solution as explained in subsection~\ref{JordanDecomposition}.
It is not a unique solution, because we may add a right eigenvector $\bm{{b}}_0$ of
arbitrary length to $\bm{a}_{0}$, and still have a solution.

\section{Construction of the resolvent}

For a real independent variable $x$ an integral expression for a solution
of the inhomogenous equation can be written down if $n$ independent
solutions $\left\{ y_k(x) \,\vert\, k=1,\ldots n\right\}$ of 
the homogeneous equation are known.
We first find a solution $G(x,x')$ of
\begin{equation}
  \left [p_{n}(x)\frac{d^{n}}{dx^{n}} + 
  p_{n-1}(x)\frac{d^{n-1}}{dx^{n-1}} + 
  \cdots + p_{0}(x)\right ]G(x,x') = \delta(x-x').
\label{GreenFunctionEquation}
\end{equation}
A solution of equation (\ref{Linearequation}) can then be expressed as
\begin{equation}
    f(x) = \int G(x, x') g(x') dx'.
\end{equation}
Since $G(x, x')$ is a solution of the homogeneous equation when $x\ne x'$ we must have
\begin{equation}
    G(x,x')  = \sum^n_{k=1} c_k y_k(x),
\end{equation}
where the coefficients $c_k$ depend on $x'$, and is different for $x > x'$ 
(denoted $c^{>}_k$) and $x<x'$ (denoted $c^{<}_k$).
Only the difference $\Delta c_k = c^{>}_k - c^{<}_k$ contributes
to the inhomogeneous solution.
The function $G(x,x')$ and its first $n-2$ derivatives with respect to $x$ 
(denoted $G^{(k)}(x,x')$) must be continuous at $x=x'$. Further,
by dividing equation~(\ref{GreenFunctionEquation}) by $p_{n}(x)$, and integrating it 
from  $x=x'-\epsilon$ to $x=x'+\epsilon$ we obtain one more condition. Altogether
\begin{align}
    &\lim_{\epsilon \to 0^+} \left[G^{(k)}(x'+\epsilon,x') - 
      G^{(k)}(x'-\epsilon,x')\right] = 0,\quad\text{for }k=0,\cdots,n-2\nonumber,\\
    \text{and}&\\
    &\lim_{\epsilon \to 0^+} \left[G^{(n-1)}(x'+\epsilon,x') - 
      G^{(n-1)}(x'-\epsilon,x')\right] = {p_n(x')^{-1}}.\nonumber
\end{align}
This can be written in matrix form,
\begin{equation}
   \begin{pmatrix}
     y_1 & y_2 &\cdots &y_n\\[0.5ex]
     y^{(1)}_1 & y^{(1)}_2 &\cdots &y^{(1)}_n\\
     \vdots&\vdots&&\vdots\\
     y^{(n-1)}_1 & y^{(n-1)}_2 &\cdots &y^{(n-1)}_n
   \end{pmatrix}
   \begin{pmatrix}
     \Delta c_1\\
     \Delta c_2\\
     \vdots\\
     \Delta c_n
   \end{pmatrix} 
   = 
   \begin{pmatrix}
     0\\
     0\\
     \vdots\\
     p_n^{-1}
   \end{pmatrix},
\end{equation}
where all quantities are evaluated at the point $x'$.
According to Cramer's rule the solution of this equation
can be expressed in terms of determinants. For $n=2$ one finds
\begin{equation}
    \begin{pmatrix}
      \Delta c_1\\
      \Delta c_2 
    \end{pmatrix}
    =
    \frac{1}{W(y_1,y_2)(x')\, p_2(x')}
    \begin{pmatrix}
      -y_2(x')\\y_1(x')
    \end{pmatrix},
\end{equation}
where $W(y_1,y_2)(x') \equiv [y_1(x') y'_2(x') - y'_1(x') y_2(x') ]$
is the Wronski determinant.
By choosing $c^{>}_1 = c^{<}_2 = 0$ this gives for $n=2$,
\begin{equation}
     G(x,x') = \frac{y_1(x_{<})\,y_2(x_{>})}{W(y_1,y_2)(x')\, p_2(x')},
\end{equation}
where $x_{<} \equiv \min(x, x')$ and $x_{>} \equiv \max(x, x')$.

$G(x,x')$ is the kernel of the integral operator ${\cal L}^{-1}$.
This method is often referred to as {\em variation of parameters\/}.
It can be extended from equations formulated on the real line to
equations formulated on well-behaved curves in
the complex plane.

For equations formulated in regions of
the complex plane one should instead search for a solution
to the problem
\begin{equation}
     {\cal L}\, \tilde{G}(z,z') = \frac{1}{(z'-z)},
\end{equation}
in terms of which
\begin{equation}
    f(z) = \int_{\cal C}\,\frac{dz'}{2\pi \text{i}}\, \tilde{G}(z,z') g(z'),
\end{equation}
will solve ${\cal L} f(z) = g(z)$ when ${\cal C}$ is a suitable curve
in the $z'$-plane, encircling the point $z$ once in the positive (anticlockwise) direction.

\chapter{The WKB approximation}

It is almost 100 years since Niels Bohr made his first formulations of a quantum
theory of matter \cite{{NielsBohrAug1913},{NielsBohrSep1913}}.
Subsequent developments by him and others, including
William Wilson \cite{Wilson1915} and Arnold Sommerfelt \cite{Sommerfeld1916}, completed
the ``old quantum theory'', and led to the formulation of the Bohr-Sommerfelt or
Sommerfelt-Wilson or Bohr-Sommerfelt-Wilson quantization rules.
With the advent of quantum mechanics and the Schr{\"o}dinger
equation these rules can be derived by the
WKB approximation \cite{Jeffreys, Wentzel, Kramers, Brillouin} to some extent.

In this thesis we have used the WKB method for many purposes:
(i)~To make {\em a priori\/} estimates of the wave-functions
before computing normalization integrals, as done in
Paper II \cite{NormalizedEigenfunctions},
(ii)~to estimate the magnitude of coefficients in the Frobenius series by
use of the leading order WKB approximation, as done in 
Paper III \cite{SeriesSolutions},
Paper IV(a) \cite{WKBLegendre}
and Paper IV(b) \cite{GeneratingFrobeniusSeries} and (iii)~to compare the
very-high-precision numerical solutions against higher order WKB results,
as done in Paper V \cite{LoopExpansion}.

The time-independent Schr{\"o}dinger equation usually has the form
\begin{equation}
  \epsilon^{2} \psi^{''}(x)=Q(x)\psi(x),\label{TimeIndependentSchrodinger}
\end{equation}
with\footnote{Note that this notation differs from the corresponding one
in paper III \cite{SeriesSolutions}, paper IV(a) \cite{WKBLegendre},
and paper IV(b) \cite{GeneratingFrobeniusSeries}.} $Q(x) = V(x)-E$.
We will mostly consider cases with the boundary condition $\lim_{x\to\pm\infty} \psi(x)=0$,
and where $V(x)$ is a polynomial in $x$. 

The standard WKB formulas can be found in most textbooks on
Quantum Mechanics, f.i.~\cite{Schiff, Kroemer}. A more thorough
discussion is given by Bender and Orszag~\cite{BenderOrszag}.
Then two leading order WKB solutions
of equation~(\ref{TimeIndependentSchrodinger}) are
\begin{equation}
  \psi_{\pm}(x) = Q(x)^{-1/4}\,
  \exp{\left(\pm\frac{\text{1}}{\epsilon}\int^x \sqrt{Q(t)}\, \text{d}t \right)}.
  \label{LeadingOrderWKB}
\end{equation}
To satisfy the boundary conditions to this order the quantization condition 
\begin{equation}
    \frac{1}{\epsilon} \int_{x_-}^{x_+}  \sqrt{-Q(x)}\, \text{d}x = \left(N + \frac{1}{2}\right)\pi,
  \label{TwoTurningPointCondition}
\end{equation}
must be fulfilled in the case of a situation with two turning points $x_\pm$ 
at which $Q(x_{\pm})=0$, so that $Q(x) \le 0$ for $x_{-} \le x \le x_{+}$.
Here $N = 0, 1, \cdots$.

There are two extensions of these formulas which may be less known.
These are (i)~the Langer correction to (\ref{LeadingOrderWKB}) and (\ref{TwoTurningPointCondition})
near a (regular) singular point, and (ii)~a more general quantization condition derived by
Dunham~\cite{Dunham} which makes higher order WKB corrections to (\ref{TwoTurningPointCondition}) quite
straightforward to compute. These extensions are discussed in this chapter.

\section{The Langer correction}

Consider a generalization of equation~(\ref{TimeIndependentSchrodinger}),
\begin{equation}
     \epsilon^2 \!\left(\frac{d^2}{dx^2}  + 
     \frac{1\!-\!\nu_+ \!-\! \nu_-}{x}\frac{d}{dx} + 
     \frac{\nu_+ \nu_-}{x^2} \right)\psi(x) = Q(x)\,\psi(x).
   \label{ODE}
\end{equation}
Such equations may f.i.~arise as radial equations of rotation
symmetric problems. The difference is that we now have a boundary
condition at $x=0$. By writing $x=\text{e}^u$ the point $x=0$ is
transformed to $u = -\infty$. Equation~(\ref{ODE}) becomes,
with $\psi(x)= \text{e}^{\frac{1}{2}(\nu_+ + \nu_-)u}\,\Psi(u)$,
\begin{equation}
     \epsilon^2 \!\left[\frac{d^2}{du^2}  - 
       \frac{1}{4}\left({\nu_+-\nu_-}\right)^2\right]\Psi(u) = 
   \text{e}^{2u}\,Q(\text{e}^{u})\,\Psi(u).
   \label{ODE_modified}
\end{equation}
This equation can now be solved by the WKB method,
and transformed back to the $x$-variable. The results are that
\begin{align}
  \psi_{\pm}(x) \approx
  x^{\nu_{\pm}} \left(\tilde{Q}(0)/\tilde{Q}(x)\right)^{1/4}\;
  \exp\left(\pm\frac{1}{\epsilon}\int_0^x
    \left[\sqrt{\tilde{Q}(t)} - \sqrt{\tilde{Q}(0)}\right] \frac{\text{d}t}{t}\right). 
  \label{LangerCorrectedWKBApproximation}
\end{align}
Here $\tilde{Q}(x) = \frac{1}{4}\epsilon^2 (\nu_+-\nu_-)^2 + x^2\,Q(x)$.
The normalization has been chosen so that 
\begin{equation*}
  \psi_{\pm}(x) \sim x^{\nu_{\pm}}\quad\text{as }x\to 0^+.
\end{equation*}
The Langer corrected quantization condition becomes
\begin{equation}
    \frac{1}{\epsilon}\int_{x_-}^{x_+} \sqrt{-\tilde{Q}(t)}\,\frac{\text{d}t}{t} = 
    \left(N+\frac{1}{2}\right)\pi,
    \label{LangerCorrectedQuantizationCondition}
\end{equation}
where $\tilde{Q}(x_{\pm})=0$ and $\tilde{Q}(x_{\pm})\le 0$ for $x_- \le x \le x_+$.
This corresponds to the classically allowed region, 
with $x_\pm$ being the classical turning points.
For the radial Schr{\"o}dinger equation in 3 dimensions,
\begin{equation}
    -\epsilon^2\left(\frac{d^2}{d r^2} + \frac{2}{r}\frac{d}{dr} 
      - \frac{\ell(\ell +1)}{r^2} \right) \psi(r) + V(r)\,\psi(r) 
    = E\,\psi(r), 
\end{equation}
with $\epsilon^2 = \hbar^2/2m$, we have $\nu_+=\ell$, $\nu_-=-(\ell+1)$. 
The quantization condition becomes
\begin{equation}
    \frac{1}{\epsilon}\int_{r_-}^{r_+} 
    \sqrt{E - V(r) - \epsilon^2\,(\ell + \text{\footnotesize $\frac{1}{2}$})^2\, r^{-2}}\;\text{d}r 
    = (N+\frac{1}{2})\pi.
\end{equation} 
In this case the Langer correction is a modification of the ``centrifugal potential'',
\begin{equation*}
  \frac{\ell(\ell+1)}{r^{2}} \longrightarrow 
  \frac{\left( \ell + \frac{1}{2}\right)^2}{ r^{2}}.
\end{equation*}
With this correction the WKB quantization formulas for the hydrogen atom
and the 3-dimensional rotation symmetric harmonic oscillator turns
out to be exact.

\section{Higher order WKB quantization condition}

\subsection{Recursive calculation of higher order corrections}

Higher order corrections to the WKB approximated wave-function, 
\begin{equation}
  \psi(x)= \exp \left[\frac{1}{\epsilon}\sum_{n=0}^{\infty}\epsilon^{n}S_{n}(x)\right].
  \label{WKBapp}
\end{equation}
can be found by substituting (\ref{WKBapp}) into (\ref{TimeIndependentSchrodinger}),
and comparing terms order-by-order in $\epsilon$. We have the equations
\begin{subequations}
  \label{WKB_series_equations}
  \begin{alignat}{1}
    &S'_0{}^2 = Q,\\
    &S''_{n-1}  + \sum_{j=0}^n S'_j S'_{n-j}  = 0,\quad\text{for $n=1, 2, \ldots$.}
    \end{alignat}
\end{subequations}
We find recursively,
\begin{subequations}
  \label{SpEquations}
  \begin{alignat}{1}
  &S'_{0} = -\sqrt{Q},\label{initialization}\\
  &S'_{n} = \frac{1}{2\sqrt{Q}}\left( S''_{n-1} + 
    \sum_{j=1}^{n-1} S'_{j}  S'_{n-j} \right).  \label{recursion}
  \end{alignat}
\end{subequations} 
The first terms of the recursion (\ref{recursion}) are 
\begin{subequations}
  \label{SpRecursion}
  \begin{alignat}{1}
  S'_{1}(x) &= -\frac{1}{4}\frac{Q^{'}(x)}{Q(x)} = 
  -\frac{1}{2} \frac{d}{dx} \log S_0(x),\label{FirstCal}\\
  S'_{2}(x) &= \frac{5}{32}\frac{Q'(x)^2}{[Q(x)]^{5/2}} - \frac{1}{8}\frac{Q''(x)}{[Q(x)]^{3/2}},
  \label{SecondCal}\\
  S'_{3}(x) &= -\frac{15}{64}\frac{{Q'(x)}^3}{Q(x)^{4}} + \frac{9}{32} \frac{Q'(x)Q''(x)}{Q(x)^{3}}
  - \frac{1}{16} \frac{Q'''(x)}{Q(x)^{2}} = -\frac{1}{2}\frac{d}{dx} \frac{S'_2(x)}{S'_0(x)}.\label{ThirdCal}
  \end{alignat}
\end{subequations}
There is another solution obtained by changing the sign of all even
terms $S'_{2m}$. 

The solutions can also be expressed in terms of multivariate polynomials.
Define the infinite-dimensional vector 
\begin{equation}
    \bm{Q} = (Q_0, Q_1,\cdots, Q_k, \cdots) \equiv 
    \left(Q(x), Q'(x), \cdots, Q^{(k)}(x),\cdots \right).
\end{equation}
Then the general term can be written as
\begin{equation}
    S'_n  = \frac{1}{2}\left[4 Q_0\right]^{(1-3n)/2}\,{\cal T}_n(\bm{Q}),
\end{equation}
where ${\cal T}_n(\bm{Q})$ is a homogeneous $n$'th order
polynomial in the components of $\bm{Q}$, with integer coefficients, and also homogeneous of
$n$'th order in derivatives. I.e., it consists of all monomials of
the form
\begin{equation*}
     Q_0^{n_0} Q_1^{n_1}\cdots Q_k^{n_k}\cdots,\quad\text{with all }n_k \ge 0,
     \quad\sum_k n_k = n,\quad\sum_k k n_k = n.
\end{equation*}
This means that number of terms in ${\cal T}_n$ is equal to the number of
partitions of $n$. The first terms are
\begin{subequations}
  \begin{alignat}{1}
     {\cal T}_0 &= -1,\\
     {\cal T}_1 &= -2\,Q_1,\\
     {\cal T}_2 &= 10\,Q^2_1 -8\,Q_0 Q_2,\\
     {\cal T}_3 &= -120\,Q^3_1 +144\,Q_0 Q_1 Q_2 -32\,Q^2_0 Q_3,\\
     {\cal T}_4 &= 2210\,Q^4_1 -3536\,Q_0 Q^2_1 Q_2 +608\,Q^2_0 Q^2_2  + 896\,Q^2_0 Q_1 Q_3 -128\,Q^3_0 Q_4.
     \end{alignat}
\end{subequations}

Equation~(\ref{SpRecursion}) indicate that each of the odd terms can be written as the derivative
of expressions involving the even terms, and hence can be integrated explicitly. This is the case in general. 
In the classically allowed region, $E \ge V(x)$ or
$Q(x) \ge 0 $, all even terms $S'_{2m}$ are imaginary and all odd terms
$S'_{2m+1}$ are real. I.e, one can write the sum $S' = \sum^\infty_{n=0} \epsilon^n S'_n$
as $S' = S'_{\text{R}} + \text{i} S'_{\text{I}}$, with $S'_R =
\sum_{m=0} \epsilon^{2m+1} S'_{2m+1}$ and $S'_{I} = \sum^\infty_{m=0}
\epsilon^{2m} S'_{2m}$ both real. They satisfy the equation
\begin{equation}
   \left( S'_{\text{R}} + \text{i} S'_{I}\right)^2 + \epsilon \left(
     S''_{\text{R}} + \text{i} S''_{\text{I}}\right) = -Q.
\end{equation}
From the imaginary part of this equation we find
\begin{equation}
    S'_{\text{R}} = -\frac{\epsilon}{2} \frac{d}{dx} \log S'_{\text{I}} = 
   -\frac{\epsilon}{2} \frac{d}{dx} \log S'_0 
   -\frac{\epsilon}{2} \frac{d}{dx} \log \left(1 + \sum^\infty_{m=1} \epsilon^{2m} \frac{S'_{2m}}{S'_0}\right),
\end{equation}
which implies that (for $m>0$)
\begin{equation}
     S'_{2m+1} = -\frac{1}{2} \frac{d}{dx} \left.\log\left( 1 +
       \sum_{k=1}^{\infty} \epsilon^{2k}
       \frac{S'_{2k}}{S'_0}\right)\right|_{\text{Order $\epsilon^{2m}$
       coefficient}}.
\end{equation}
Hence $S'_{2m+1}$ is the derivative of a single-valued function when $m>0$. The next example beyond
equation~(\ref{ThirdCal}) is 
\begin{align}
    S'_5 = -\frac{1}{2}\frac{d}{dx}\left(\frac{S'_4}{S'_0} - \frac{1}{2}\frac{{S'_2}^2}{{S'_0}^2}\right).
\end{align}
This relation is straightforward to verify with a computer algebra program,
but the explicit expressions in terms of $Q$ are too lengthy to write down.

One should be aware that the expansion (\ref{WKBapp})  will not converge towards
the exact result in general. Consider a case with a non-constant $Q<0$ everywhere,
so that the WKB solution describes
a wave moving to (say) the right. The higher order corrections will modify the shape
of this right-moving wave, but never generate a left-moving wave. However, the exact solution
for a quantum particle moving over a potential barrier will usually contain an exponentially
small back-scattered wave. Hence one should expect exponentially small corrections to the
WKB-series considered above.

\subsection{The Dunham formula}

We now return to the two-turning point eigenvalue problem 
(\ref{TimeIndependentSchrodinger}), with a potential $V(x)$
which is assumed to be analytic in $x$. 
The two-turning point  quantization condition (\ref{TwoTurningPointCondition}) has been
generalized to arbitrary order in $\epsilon$ by Dunham~\cite{Dunham}
(apparently as part of a Ph.D thesis at Harvard, after which no published research
by the author seems to exist),
\begin{equation}
  \frac{1}{2\text{i}\epsilon}\oint \sum_{n=0}^{\infty} \epsilon^n\,S_{n}^{'}(z) \text{d}z = N\pi. \label{exactquantization}
\end{equation}
The above integral is a complex contour  integral which encircles a branch cut between the two classical turning
points $x_{\pm}$ on the real axis. The WKB expansion breaks down near the turning points, but by
extending the expansion into the complex plane the turning points can be avoided. The quantization condition
is obtained by requiring the wave-function to be single valued.
The integral in (\ref{exactquantization}) is finite because the  contour {\em encircles\/} the turning points
instead of passing through them. The quantization condition (\ref{TwoTurningPointCondition}) is recovered
by considering the first two terms of the expansion. The contribution from $S_0$ becomes
\begin{equation*}
  \frac{1}{2 \text{i}\epsilon}\oint S_{0}^{'}(z) \text{d}z = -\frac{1}{2 \text{i}\epsilon}\oint \sqrt{Q(z)} \text{d}z
  =\frac{1}{\epsilon}\int^{x_2}_{x_1}\sqrt{-Q(x)} \text{d}x,
\end{equation*}
and the contribution from $S_1$ becomes
\begin{equation*}
  \frac{1}{2 \text{i}}\oint S_{1}^{'}(z) \text{d}z =   -\frac{1}{8\text{i}}\oint \frac{d}{dz}\log Q(z) \text{d}z
  = -\frac{1}{8\text{i}}\, 4\pi \text{i} = -\frac{\pi}{2},
\end{equation*}
since the logarithmic integral encircles two simple zeros. Hence the Dunham quantization condition becomes
\begin{equation}
     \frac{1}{\epsilon}\int^{x_2}_{x_1}\sqrt{-Q(x)} \text{d}x + 
     \frac{1}{2\text{i}\epsilon}\oint \sum_{m=1}^{\infty} \epsilon^{2m}\,S_{2m}^{'}(z) \text{d}z = \left(N+\frac{1}{2} \right)\pi.
     \label{DunhamQuantizationCondition}
\end{equation}

\subsection{Exactly solved cases}

The quantization condition~(\ref{DunhamQuantizationCondition}) seems to depend
only on $Q(z)$ in the region near the branch cut from $x_1$ to $x_2$, and therefore
obviously cannot always be correct. We could modify the potential in a far-away
region, thereby changing the exact eigenvalues, without changing the value of
$Q(z)$ in the region of integration. However, this is only possible with a
non-analytic potential. The interesting question is
whether~(\ref{DunhamQuantizationCondition}) is exact or not for \emph{analytic} potentials.
To our knowledge this had proven to be true for all cases where the expansion
in~(\ref{DunhamQuantizationCondition}) can be carried out to all orders,
and a comparison with exactly known solutions can be
made \cite{Dunham, Bailey1964, RosenzweigKrieger, Bender_etal} .

This is known to be the case for
\begin{enumerate}

\item 
the harmonic oscillator, 
\begin{equation*}
   Q(z) = z^2 - E,
\end{equation*}
where the integration contour of all correction terms in~(\ref{DunhamQuantizationCondition})
can be deformed to infinity, leading to the conclusion that all correction
terms vanish. Hence, the condition~(\ref{DunhamQuantizationCondition})
reduces to the standard first order WKB results, which is known to
reproduce the correct result.

\item
the Morse potential \cite{MorsePotential},
\begin{equation*}
  Q(z) = \text{e}^{-2z} - \text{e}^{-z} -E,
\end{equation*}
where the same procedure can be carried out after
a change integration variable, $u = \text{e}^{-z}$, 
in~(\ref{DunhamQuantizationCondition}).

\item
the radial equation of the hydrogen atom, 
\begin{equation*}
  Q(z) = a\,z^{-1}+ b\,z^{-2} -E,
\end{equation*}
where the same procedure can also be carried out after a change of
integration variable, $u = z^{-1}$,
in~(\ref{DunhamQuantizationCondition}).

\item
the radial equation of the rotation symmetric harmonic oscillator,
\begin{equation*}
 Q(z) = a\,z^2 + b\,z^{-2} - E.
\end{equation*}
One may show that only a subset of the terms in the WKB expansion
have a non-zero integral. These terms can be computed explicitly.

\item
the P{\"o}schl-Teller potential~\cite{PoschlTellerPotential}, 
\begin{equation*}
   Q(z) = a\,\text{sech}^2 z - E.
\end{equation*}
By introducing the parameter $u=\sinh z$ one may show
that only a subset of the terms in the WKB expansion have
a non-zero integral. These terms can be computed explicity \cite{Bender_etal}.

\end{enumerate}
All cases have the common property that they have only a single branch cut
in the full complex plane of the final integration variable.

\subsection{Polynomial potentials}

Now restrict to the case that $Q(x)$ is a polynomial of order $K$. We observe from (\ref{FirstCal}-\ref{ThirdCal})
that $S_{n}^{'}$  has the form
\begin{equation}
  S_{n}^{'}(x) = \frac{1}{[Q(x)]^{(3n-1)/2}} P_{n}(x),
\end{equation}
where $P_n$ is a polynomial of order $(K-1)n$.
By substituting this ansatz into (\ref{recursion}) we verify that it is correct, and find the recursion
relation
\begin{equation}
  P_{n+1} = \frac{1}{4}(3n-1)Q'\,P_{n} -\frac{1}{2}Q\,P'_{n} + 
  \frac{1}{2}\sum_{j=1}^{n}P_{j}\,P_{n+1-j},
\end{equation}
with $P_0=-1$. This gives 
\begin{align}
   P_1 &= \frac{1}{4}Q',\\ 
  P_2 &= \frac{5}{32} {Q'}^2 -\frac{1}{8}Q Q'',\\
  P_3 &= \frac{15}{64} {Q'}^3 - \frac{9}{32} Q Q' Q'' + \frac{1}{16} Q^2\,Q''',
\end{align}
and so on. Since every odd $S'_{2m+1}, (m=1, 2, \cdots)$ is the derivative of a single-valued function it
does not contribute to the quantization condition (\ref{exactquantization}).
Thus, (\ref{exactquantization}) simplifies to a sum over even-numbered terms
only, 
\begin{equation}
   \frac{1}{2 \text{i} \epsilon}\oint \sum_{m=0}^{\infty} \epsilon^{2m} S_{2m}^{'}(z) dz = (N+\frac{1}{2})\pi.
\end{equation}
Further, we may subtract any total derivative of the form
\begin{equation}
    \frac{d}{dx} \frac{R_{2m}(x)}{\left[ Q(x)\right]^{\alpha_m}} = \frac{1}{[Q(x)]^{\alpha_m+1}}
    \left[Q(x) R'_{2m}(x) - \alpha_m\,Q'(x)\,R_{2m}(x)\right]
\end{equation}
(with $R_{2m}$ a single-valued function) from $S_{2m}$ without changing the value of the contour integral.
By choosing $\alpha_m = 3m-\frac{3}{2}$, and $R_{2m}$ a polynomial of order $(K-1)(2m-1)$, this can
be used to replace $P_{2m}$, of order $(K-1)2m$,  by a polynomial $\tilde{P}_{2m}$ of order $(K-2)$ or less. 
% except possibly for a term $P_{2m,r}\,x^r$  where $r+1 = (3m-\frac{1}{2})K$ is an integer.

\part{Appendices}

\chapter{Differential Equations}

In this appendix we provide a fairly complete analysis of
the general expansion of solutions to a general first
order homogeneous linear matrix differential equations at
a regular singular point, for the cases when higher order
logarithmic terms occur. The derivations are done for the
purpose of later numerical implementations, but we have not
yet done such implementations.

\section{Matrices with fewer eigenvectors than eigenvalues\label{JordanDecomposition}}

An arbitrary $D\times D$ matrix ${\cal K}$ cannot be completely diagonalized
in general. Although the polynomial eigenvalue equation,
\begin{equation}
     \det \left( {\cal K} - \mu \right) = 0,
\end{equation}
always has $D$ solutions $\left\{ \mu_d \,\vert\, d=1,\ldots, D \right\}$
counting multiplicities, a $d$-fold degenerate distinct root $\mu$ may have
$d_{\mu} < d$ linearly independent eigenvectors (with $d_{\mu} \ge 1$). I.e.,
the {\em geometric multiplicity\/} $d_{\mu}$ of an eigenvalue $\mu$
may be lower than its {\em algebraic multiplicity\/} $d$. 

However, any $D\times D$ matrix ${\cal K}$  can be brought to {\em Jordan normal form\/} \cite{SergeLang}.
I.e., if ${\cal K}$ has $d$ linearly independent eigenvectors it can be similarity
transformed to a block diagonal form,
\begin{equation}
    {\cal J} = {\cal S}^{-1}\, {\cal K}\,{\cal S} = 
    \begin{pmatrix}
        \bm{J}_1&\cdots&\bm{0}\\\vdots&\ddots&\vdots\\\bm{0}&\cdots&\bm{J}_d
     \end{pmatrix},
     \label{SimilarityTransformation}
\end{equation}
where each block $\bm{J}_n$ is a $d_n \times d_n$ bidiagonal matrix of form
\begin{equation}
     \bm{J}_n =
     \begin{pmatrix}
      \mu_n&1&\cdots&0\\
      0&\mu_n&\ddots&0\\
      \vdots&\vdots&\ddots&\vdots\\
      0&0&\cdots&\mu_n
     \end{pmatrix}.
\end{equation}
There may be more than one block for each distinct eigenvalue.
Now observe that there is a sequence of column vectors,
\begin{equation*}
    \bm{e}^{(k)} = (0,\cdots,\underbrace{1}_{\text{position } k},\cdots,0)^T, \quad k=1,\cdots,d_n,
\end{equation*}
(where ${}^T$ stands for transposition) such that
\begin{align}
    \left(\bm{J}_n-\mu_n\right)\,\bm{e}^{(k)} &= \bm{e}^{(k-1)},\quad\text{for }k=2,\cdots,d_n\\
    \left(\bm{J}_n-\mu_n\right)\,\bm{e}^{(1)} &=  0.\label{RightEigenvalueEquation}
\end{align}
There is also a corresponding sequence of row vectors,
\begin{equation*}
    \bm{\bar{e}}^{(k)} = (0,\cdots,\underbrace{1}_{\text{position } d_n+1-k},\cdots,0), \quad k=1,\ldots,d_n,
\end{equation*}
such that
\begin{align}
    \bm{\bar{e}}^{(k)}\,\left(\bm{J}_n-\mu_n\right)\, &= \bm{\bar{e}}^{(k-1)},\quad\text{for }k=2,\cdots,d_n\\
    \bm{\bar{e}}^{(1)}\,\left(\bm{J}_n-\mu_n\right)\, &=  0.\label{LeftEigenvalueEquation}
\end{align}

Equation~(\ref{RightEigenvalueEquation}) means that $\bm{e}^{(1)}$ is a right eigenvector of $\bm{J}_n$,
while equation~(\ref{LeftEigenvalueEquation}) means that $\bm{\bar{e}}^{(1)}$ is a left eigenvector of $\bm{J}_n$,
both with eigenvalue $\mu_n$. There is a (perhaps unusual) orthonormality relation for these vectors.
Define the ``backward identity matrix'' $\bm{E}$ as
\begin{equation}
  \bm{E} = 
  \begin{pmatrix}0&&1\\&\adots&\\1&&0
  \end{pmatrix}.
\end{equation}
Then we have the relation
\begin{equation}
   \bm{\bar{e}}^{(k)}\, \bm{E} \bm{e}^{(\ell)} = \delta_{k\ell}.
\end{equation}

One may extend the column vectors $\bm{e}^{(k)}$ to
$D$-dimensional column vectors $\bm{\varepsilon}^{(k,n)}$, and  
the row vectors $\bm{\bar{e}}^{(k)}$ to 
$D$-dimensional row vectors $\bm{\bar{\varepsilon}}^{(k,n)}$ by inserting
them into the $n$'th block, with zeros in all other blocks. They satisfy
the relations
\begin{align}
    \left({\cal J} -\mu_n\right)\,\bm{\varepsilon}^{(k,n)} &= \bm{\varepsilon}^{(k-1,n)},
    \quad\text{for }k=2,\ldots,d_n\nonumber\\[-2.5ex]
    &\label{LadderEquations}\\
    \left({\cal J} -\mu_n\right)\,\bm{\varepsilon}^{(1,n)}  &=  0,\nonumber\\
\end{align}
for $n=1,2,\cdots d$. These relations, combined
with equation~(\ref{SimilarityTransformation}), means that we can find a solution
to the equations
\begin{align}
    \left({\cal K}-\mu_n\right)\,\bm{x}^{(1)} &= 0,\nonumber\\
    \left({\cal K}-\mu_n\right)\,\bm{x}^{(2)} &= c_2\,\bm{x}^{(1)}\nonumber \\
    &\;\;\vdots\\
    \left({\cal K}-\mu_n\right)\,\bm{x}^{(d_n)} &= c_{d_n}\,\bm{x}^{(d_n-1)},\nonumber
\end{align}
in terms of the vectors ${\cal S}\bm{\varepsilon}^{(k,n)}$.
We make the ansatz $\bm{x}^{(k)} = N_k\,{\cal S}\bm{\varepsilon}^{(k,n)}
+ P_k\,{\cal S}\bm{\varepsilon}^{(1,n)}$, and find
\begin{align}
     \bm{x}^{(1)} &= N_1\,{\cal S}\bm{\varepsilon}^{(1,n)},\nonumber\\[-2.5ex]
     \\
     \bm{x}^{(k)} &= N_1\, \prod_{j=2}^k\, c_j\,{\cal S}\bm{\varepsilon}^{(k,n)},
     \quad\text{for }k=2,\ldots,d_n,\nonumber
\end{align}
where $N_1$ is a free parameter.

\section{General case}

In the general case the initialization step constitutes of finding
a complete set of solutions to the equation
\begin{equation}
    \left(p_1 z \frac{d}{dz} - \bm{A}_0  \right) \bm{a}(z) = 0.
\end{equation}

\subsection{The Jordan normal form}
$\bm{A}_0$ can be brought to Jordan normal form by a
similarity trans\-form \cite{SergeLang},
\begin{equation}
     \bm{A}_0 = {\cal S} {\cal J} {\cal S}^{-1},
\end{equation}
with ${\cal J}$ block diagonal
\begin{equation}
    {\cal J} = 
    \begin{pmatrix}
        \bm{J}_1&\cdots&\bm{0}\\\vdots&\ddots&\vdots\\\bm{0}&\cdots&\bm{J}_d
     \end{pmatrix}.
     \label{JordanNormalForm}
\end{equation}
Here each block $\bm{J}_n$ is a $d_n \times d_n$ bidiagonal matrix of form
\begin{equation}
     \bm{J}_n =
     \begin{pmatrix}
      \lambda_n&1&\cdots&0\\
      0&\lambda_n&\ddots&0\\
      \vdots&\vdots&\ddots&1\\
      0&0&\cdots&\lambda_n
     \end{pmatrix}.
\end{equation}
This representation is unique up to permutation of the blocks.
There may be more than one block for each distinct eigenvalue.
Introduce
\(
    \bm{K}_n \equiv \bm{J}_n - \lambda_n
\),
and observe that (we let ${\ }^T$ denote transposition)
\begin{align}
   \bm{K}_n^T  
   \begin{pmatrix}
       \alpha_1\\\alpha_2\\ \vdots\\\alpha_{d_n} 
   \end{pmatrix} =    
   \begin{pmatrix}
       0\\\alpha_1\\\vdots\\\alpha_{d_n\!-\!1} 
   \end{pmatrix},\quad
   \bm{K}_n  
   \begin{pmatrix}
       \alpha_1\\\alpha_2\\ \vdots\\\alpha_{d_n} 
   \end{pmatrix} =    
   \begin{pmatrix}
       \alpha_2\\\vdots\\\alpha_{d_n}\\0 
   \end{pmatrix}
\end{align}
This means that the matrix
\begin{equation}
  \bm{K}_n \bm{K}_n^T = \text{diag}(1,1,\ldots,1,0)
\end{equation}
is a projection onto the subspace orthogonal to
the left eigenvector $\bm{\bar{e}}_n =  (0,\ldots,0,1)$ of $\bm{K}_n$.
For a coordinate invariant description, one notes by
taking the scalar product of the equation
\begin{equation}
    \bm{K}_n \bm{f} = \bm{g}
    \label{SingularEquation}
\end{equation}
with the left eigenvector $\bm{\bar{e}}_n$, that
\begin{equation*}
    \bm{\bar{e}}_n \bm{K}_n \bm{f} = 0 = \bm{\bar{e}}_n\cdot \bm{g}, 
\end{equation*}
since $\bm{\bar{e}}_n \bm{K}_n = \bm{0}$. Hence, equation~(\ref{SingularEquation})
has a solution only if the solubility condition
\begin{equation}
   \bm{\bar{e}}_n\cdot \bm{g} = 0
   \label{ConditionForSolution}
\end{equation}
is fulfilled. Further, when there is
a solution $\bm{f}$ it cannot be unique; we may always add a
term $\alpha\, \bm{e}_n$ to $\bm{f}$, since $\bm{K}_n\, \bm{e}_n = \bm{0}$.
When equation~(\ref{ConditionForSolution}) holds the general solution is
\begin{equation}
   \bm{f} =  \bm{K}^T_n\,  \bm{g} + \alpha\,\bm{{e}}_n,
\end{equation}
where the coefficient $\alpha$ can be chosen freely.

\subsection{Solution of a homogeneous block equation}

The initialization step consists of finding the general solution of the homogeneous equation
\begin{equation}
    \left(p_1 z \frac{d}{dz} - \bm{J}_n\right) \bm{\varphi}(z) = 0.
    \label{BlockHomogeneousEquation}
\end{equation}
We make the ansatz
\begin{equation}
    \bm{\varphi}(z) = \sum^{d_n-1}_{k=0} \bm{\varphi}_k\, z^\nu\, \log(z)^{d_n-1-k},
\end{equation}
where $p_1 \nu = \lambda_n$ is the eigenvalue of $\bm{A}_0$ corresponding to the $n$'th block.
Inserted into equation~(\ref{BlockHomogeneousEquation}) this leads to the conditions
\begin{equation}
  \bm{K}_n \bm{\varphi}_k =  p_1 \left(d_n \!-\!k \right)\,\bm{\varphi}_{k-1},
  \quad\text{for } k=0,\ldots,d_n\!\!-\!\!1,
\end{equation}
where $\bm{\varphi}_{-1}\equiv \bm{0}$. This means that $\bm{\varphi}_0$ must be
an eigenvector of $\bm{K}_n$,
\begin{equation*}
    \bm{\varphi}_0  = \alpha_0\,\bm{e}_n \equiv \alpha_0\,\bm{e}^{(0)}_n
\end{equation*}
The next $d_n\!-\!1$ equations can be solved iteratively,
\begin{equation*}
    \bm{\varphi}_{k} = p_1 (d_n\!-\!k) \bm{K}^T_n \bm{\varphi}_{k-1} + \alpha_k\,\bm{e}_n 
    = \sum_{j = 0}^k  \alpha_{j}\,P^{k}_j\, \bm{e}^{(k-j)}_n,
\end{equation*}
for $k=1,\cdots,d_n\!-\!1$.
Here
\begin{equation}
      \bm{e}^{(\ell)}_n \equiv \left({\bm{K}^T_n}\right)^\ell \bm{e}_n,\quad\text{and }
      P^{k}_j \equiv \prod_{\ell=j+1}^k p_1 \left(d_n \!-\! \ell\right)\quad\text{with }
      P^{k}_{k} \equiv 1.
\end{equation}
Hence we have found a $d_n$-dimensional space of solutions to equation~(\ref{BlockHomogeneousEquation}),
\begin{equation}
     \bm{\varphi}(z) = \sum_{j=0}^{d_n\!-\!1} \alpha_j\,
     \sum^{d_n\!-\!1}_{k=j} P^k_j \,\bm{e}_n^{(k-j)}\,z^{\nu} \log(z)^{d_n-1-k} \equiv 
     \sum_{j=0}^{d_n\!-\!1} \alpha_j\,\bm{\varphi}^{(j)}(z),
\end{equation}
when $p_1 \nu$ is an eigenvalue of $\bm{J}_n$ with algebraic multiplicity $d_n$ and geometric multiplicity $1$.

\subsection{Solution of an inhomogeneous block equation (regular case)}

The recursion step essentially consists of solving an inhomogeneous equation like
\begin{equation}
    \left(p_1 z \frac{d}{dz} - \bm{J}_n\right) \bm{\varphi}(z) = \sum_{k=0}^{\ell} \bm{\chi}_k\,z^\mu \log(z)^{\ell - k}.
    \label{BlockInhomogeneousEquation}
\end{equation}
Assume that $p_1 \mu$ is not an eigenvalue of $\bm{J}_n$,
so that $p_1\mu - \bm{J}_n$ is invertible. We make the solution ansatz
\begin{equation}
       \bm{\varphi}(z) = \sum_{k=0}^{\ell}  \bm{\varphi}_k\,z^\mu \log(z)^{\ell -k}.
\end{equation}
Inserted into equation~(\ref{BlockInhomogeneousEquation}) this leads to the conditions
\begin{equation}
      (p_1\mu - \bm{ J}_n) \bm{\varphi}_k = \bm{\chi}_k - p_1\left(\ell+1-k\right)\bm{\varphi}_{k-1}\quad\text{with }\bm{\varphi}_{-1} = \bm{0}.
\end{equation}
These can be solved recursively as
\begin{equation}
       \bm{\varphi}_k = \left(p_1\mu - \bm{ J}_n \right)^{-1}\left[ \bm{\chi}_k - p_1\left(\ell+1-k\right)\bm{\varphi}_{k-1}\right].
\end{equation}

\subsection{Solution of an inhomogeneous block equation (singular case)}

\begin{figure}[H]
\begin{center}
\includegraphics[trim= 30ex 110ex 28ex 32ex, clip, width=\linewidth]{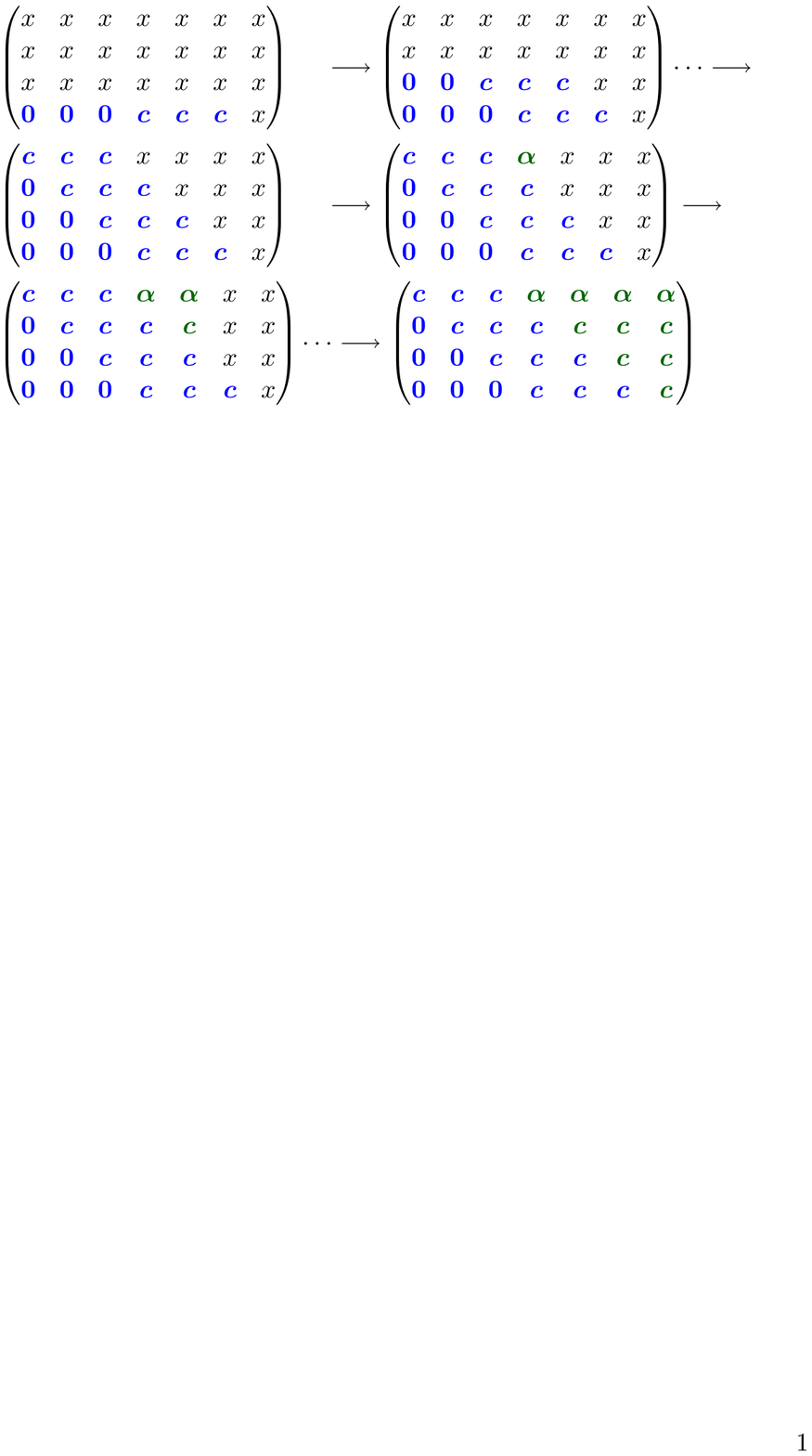}
\end{center}
\caption{\label{SolutionScheme}
This figure give a schematic description of how one can solve the
recursion equations at a step $m$, when $m +\nu_1$ is a multiple
root of the indicial equation,
and where one already have logarithmic terms in the expansion. 
}
\end{figure}

Assume that $p_1\nu = \lambda_n$ is the unique eigenvalue of $\bm{J}_n$, 
i.e.~$\left(p_1\nu - \bm{ J}_n\right) = -\bm{K}_n$. To solve
equation~(\ref{BlockInhomogeneousEquation}) we make the ansatz
\begin{equation}
      \bm{\varphi}(z) = \sum_{k=0}^{d_n + \ell} \bm{\varphi}_k\, z^\nu \log(z)^{d_n + \ell-k},
\end{equation}
which inserted leads to the condition
\begin{equation}
      \bm{K}_n \bm{\varphi}_{k+1} = -\bm{\chi}_{k+1} + p_1 \left( d_n \!+\! \ell\! \!-\! k\right) \bm{\varphi}_{k}.
      \label{SolutionCondition}
\end{equation}
Here one should interpret
$\bm{\varphi}_{-1} = \bm{0}$, and $\bm{\chi}_k = \bm{0}$ for $k=0, 1\cdots, d_n \!-\! 1$.
Since $\bm{\bar{e}}_n \bm{K}_n = 0$, equation~(\ref{SolutionCondition}) can only have a solution
if the right-hand side is  orthogonal to $\bm{\bar{e}}_n$. I.e.,
\begin{equation}
      \bm{\bar{e}}_n\cdot\bm{\chi}_{k+1} = p_1 \left(d_n \!+\! \ell \!-\! k\right)\,\bm{\bar{e}}_n\cdot\bm{\varphi}_{k}.
\end{equation}
This determines the last component of all but one of the $\bm{\varphi}_k$'s,
\begin{equation}
      \bm{\varphi}_k = C_k^{(d_n\!-\!1)}\,\bm{e}^{(d_n\!-\!1)}_n + \bm{\tilde{\varphi}}^{(1)}_k,
      \quad\text{for }k=0, 1,\cdots,d_n \!+\! \ell \!-\! 1.
      \label{SolubilityCondition}
\end{equation}
Here 
\begin{equation}
     C_k^{(d_n\!-\!1)} = \frac{(\bm{\bar{e}}_n\cdot \bm{\chi}_{k+1})}{p_1(d_n \!+\! \ell \!-\! k)} = 
     \frac{(\bm{\bar{e}}^{(0)}_n\cdot \bm{\chi}_{k+1})}{p_1(d_n \!+\! \ell \!-\! k)},
\end{equation} 
and  $\bm{\tilde{\varphi}}^{(1)}_k$ denotes the restriction of $\bm{\varphi}_k$ to the space orthogonal to $\bm{\bar{e}}_n$.
I.e.~ the first $d_n \!-\!1$ components of $\bm{\varphi}_k$. We define $\bm{\tilde{\chi}}^{(1)}_k$ in the same manner.
Insert the partial solution~(\ref{SolubilityCondition}) into equation~(\ref{SolutionCondition}), 
and use that $\bm{K}_n \bm{e}^{(d_n\!-\!1)}_n = \bm{e}^{(d_n\!-\!2)}_n$.  Since $\bm{K}_n \bm{\tilde{\varphi}}_k^{(1)}$
is also orthogonal to the space spanned by $\bm{e}^{(d_n \!-\!2)}_n$, we can use this to determine the last component
of all but two of the $\bm{\tilde{\varphi}}^{(1)}_k$'s,
\begin{equation}
      \bm{\tilde{\varphi}}_k^{(1)} = C^{(d_n \!-\! 2)}_k \bm{e}^{(d_n \!-\! 2)}_n + \bm{\tilde{\varphi}}^{(2)}_k,
       \quad\text{for }k=0,1, \cdots, d_n \!+\! \ell \!- \!2.
\end{equation}
Here
\begin{equation}
     C^{(d_n \!-\! 2)}_k = \frac{1}{p_1(d_n \!+\! \ell \!-\! k)} 
     \left[ (\bm{\bar{e}}^{(1)}_n\cdot \bm{\tilde{\chi}}_{k+1}^{(1)} ) + C^{(d_n \!-\! 1)}_{k+1} \right],
\end{equation}
and $\bm{\tilde{\varphi}}^{(2)}_k$ denotes the restriction of $\bm{\varphi}_k$ to the space orthogonal to 
$\bm{\bar{e}}^{(0)}_n$ and $\bm{\bar{e}}^{(1)}_n$.
I.e., the first $d_n \!-\!2$ components of $\bm{\varphi}_k$. We define $\bm{\tilde{\chi}}^{(2)}_k$ in the same manner.
Note that we cannot determine $C^{(d_n \!-\! 2)}_{d_n+\ell - 1}$ yet, because we do not know $C^{(d_n \!-\! 1)}_{d_n+ \ell}$.
The solution process thus far is illustrated by the first two frames of figure~\ref{SolutionScheme}, for the case that $\ell=2$
and $d_n=4$,

This process can be continued. At step $r$  the vector $\bm{K}_n \bm{\tilde{\varphi}}^{(r-1)}_k$ is orthogonal to\\
$\bm{\bar{e}}^{(0)}_n, \bm{\bar{e}}^{(1)}_n, \ldots, \bm{\bar{e}}^{(r)}_n$, leading to
\begin{equation}
     \bm{\tilde{\varphi}}_k^{(r \!-\! 1)} = C_k^{(d_n \!-\! r)}\,\bm{e}^{(d_n\!-\!r)}_n + \bm{\tilde{\varphi}}_k^{(r)},
       \quad\text{for }k=0,1, \dots, d_n \!+\! \ell \!-\! r,
\end{equation}
with
\begin{equation}
      C_k^{(d_n \!-\! r)} = \frac{1}{p_1(d_n \!+\! \ell \!-\! k)} 
     \left[ (\bm{\bar{e}}^{(r \!-\! 1)}_n\cdot \bm{\tilde{\chi}}_{k+1}^{(r \!-\! 1)} ) + C^{(d_n \!-\! r \!+\! 1)}_{k+1} \right].
\end{equation}
After $d_n$ steps all consequences of the solubility condition (\ref{SolubilityCondition})
have been deduced, as indicated by the third frame of figure~\ref{SolutionScheme}. The first $\ell+1$ vectors
$\bm{\varphi}_{0},\ldots, \bm{\varphi}_{\ell}$ are completely determined at this stage. To determine
the last $d_n$ vectors completely we solve equation~(\ref{SolutionCondition}) in the forward direction,
\begin{equation}
    \bm{\varphi}_k = -\bm{K}^T_n\, \bm{\chi}_k + 
    p_1(d_n + \ell + 1 - k)\bm{K}^T_n\, \bm{\varphi}_{k-1} + \alpha_k\, \bm{e}_n,
\end{equation}
for $k=\ell \!+\! 1,\ldots,\ell \!+\! d_n$. Each solution involves an
arbitrary constant $\alpha_k$. This process is indicated by the last three frames
in figure~\ref{SolutionScheme}.

\chapter{WKB Quantization of the Quartic Potential}

The contents of this appendix started out as a small example of the WKB quantization method
to high orders. It soon grew in magnitude and resulted in
Paper V \cite{LoopExpansion} --- and delayed the completion of my thesis with
several weeks. It provides more details of the computations reported in Paper V, for
the high order WKB analysis of the problem
\begin{equation}
    \left[-\epsilon^2 \frac{d^2}{dx^2} + x^4\right] \psi(x) = E \psi(x).
\end{equation}
One can set the parameter $\epsilon=1$ without loss of generality,
but it is useful for initial organization of the WKB expansion.

\section{High order WKB expansion}

As an example consider the case $V(x)=x^4$. The first few polynomials $P_{2m}$ and $\tilde{P}_{2m}$ become
\begin{center}
\begin{tabular}{|l|l|c|}
\hline
&&\\[-2.5ex]
$n$&\multicolumn{1}{c}{$P_{n}(x)$}&$\tilde{P}_{n}(x)$\\
\hline
&&\\[-2.5ex]
2&{\footnotesize $x^6 + \frac{3}{2}E\,x^2$}&{\footnotesize $\frac{1}{2}E\,x^2$}\\[0.4ex]
4&{\footnotesize $14 x^{12} + \frac{333}{4} E\,x^8 
+ \frac{321}{8} E^2\,x^4 + \frac{3}{4}E^3$}&{\footnotesize $-\frac{77}{1\,768} E^3$}\\[0.4ex]
6&{\footnotesize$671 x^{18} + \frac{3\,223}{4} E\,x^{14} + \frac{104\,595}{8} E^2\,x^{10} + 
\frac{63\,075}{16} E^3\,x^6 +\frac{279}{2} E^4\,x^2$}&{\footnotesize $-\frac{61\,061}{62\,928} E^4\,x^2$}\\[0.5ex]
\hline
\end{tabular}
\end{center}
The general pattern is that
\begin{align}
   \tilde{P}_{4\ell}(x) &= (-1)^{\ell} E^{3\ell}\, p^{(\text{e})}_{\ell},\label{pe}\\
   \tilde{P}_{4\ell +2}(x) &= (-1)^{\ell}\,E^{3\ell+1}\,x^2\,p^{(\text{o})}_{\ell},\label{po}
\end{align}
where the coefficients $p_{\ell}^{(\text{e})}$ and
$p_{\ell}^{(\text{o})}$ are positive rational numbers. F.i., as can be seen from the table above,
$p^{(\text{o})}_0 = \frac{1}{2}$, $p^{(\text{e})}_1 = \frac{77}{1\,768}$, $p^{(\text{o})}_1 = \frac{61\,061}{62\,928}$ .
Thus we have to do two types of integrals
\begin{align*}
    I^{\text{(e)}}_{k} &= \frac{1}{2\text{i}} \oint \frac{1}{\left(z^4 - E\right)^{k+1/2}}\,dz,\quad\text{for $k=6\ell-1$},\\
    I^{\text{(o)}}_{k}  &=  \frac{1}{2\text{i}} \oint \frac{z^2}{\left(z^4 - E\right)^{k+1/2}}\,dz,\quad\text{for $k=6\ell+2$}.
\end{align*}
We find, by writing $z= E^{1/4} u$, and deforming the integral along the real axis,
\begin{align}
     I^{\text{(e)}}_{-1} &= \frac{1}{2\text{i}} \oint {\left(z^4 - E\right)^{1/2}}\,dz = E^{3/4}\,\int^{1}_{-1} du \sqrt{1-u^4}\nonumber\\ 
     &= \frac{1}{2} B(\frac{1}{4},\frac{3}{2})\,E^{3/4} =  \frac{1}{3} B(\frac{1}{4},\frac{1}{2})\,E^{3/4},
\end{align}
where $\mathrm{B}(a, b) = \Gamma(a)\Gamma(b)/\Gamma(a+b)$ is the Beta function.
The remaining integrals of interest cannot be deformed to convergent integrals along the real axis,
but they are $E$-derivatives of integrals which can be deformed.  For $k=0$ we find, 
by writing $z= E^{1/4} u$, and deforming the integral along the real axis,
\begin{align*}
    I^{\text{(e)}}_{0} &= \frac{1}{2\text{i}} \oint \frac{1}{\left(z^4 - E\right)^{1/2}}\,dz 
    = 2 E^{-1/4} \int_0^1 \frac{1}{\sqrt{1-u^4}}\, du = 
    \frac{1}{2}\mathrm{B}(\frac{1}{4}, \frac{1}{2})\, E^{-1/4},\\
    I^{\text{(o)}}_{0}  &=  \frac{1}{2\text{i}} \oint \frac{z^2}{\left(z^4 - E\right)^{1/2}}\,dz
    = 2 E^{1/4} \int_0^1 \frac{u^2}{\sqrt{1-u^4}}\, du 
    = \frac{1}{2} \mathrm{B}(\frac{3}{4},\frac{1}{2})\,E^{1/4},
\end{align*}
By differentiating these relations $k$ times with respect to $E$ we find
\begin{align}
   I^{\text{(e)}}_k &= (-1)^{k}\frac{1}{2}\,\frac{(\frac{1}{4})(\frac{1}{4}+1)
     \cdots(\frac{1}{4}+k-1)}{(\frac{1}{2})(\frac{1}{2}+1)\cdots(\frac{1}{2}+k-1)}
   \,   \mathrm{B}(\textstyle{\frac{1}{4}},\textstyle{\frac{1}{2}})\,E^{-1/4 -k}\nonumber \\
   &= (-1)^{k} \frac{1}{2}\,\mathrm{B}(\textstyle{\frac{1}{4}},\textstyle{\frac{1}{2}-k})\,E^{-1/4 -k},\label{Ie}\\
   I^{\text{(o)}}_k &= (-1)^{k} \frac{1}{2}\,\frac{(-\frac{1}{4})(-\frac{1}{4}+1)
     \cdots(-\frac{1}{4}+k-1)}{(\frac{1}{2})(\frac{1}{2}+1)\cdots(\frac{1}{2}+k-1)}\, 
   \mathrm{B}(\textstyle{\frac{3}{4}}, \textstyle{\frac{1}{2}})\,E^{1/4-k} 
   \nonumber\\
   &= (-1)^{k} \frac{1}{2}\, \mathrm{B}({\textstyle \frac{3}{4}},{\textstyle \frac{1}{2}}-k)\,E^{1/4-k}.\label{Io}
\end{align}
We introduce the quantity
\begin{equation}
   \rho = \frac{1}{\pi^4}{\mathrm{B}({\textstyle \frac{1}{4},\frac{1}{2}})}^4 = \frac{1}{\pi^2}
   \,\frac{{\Gamma(\frac{1}{4})}^4}{{\Gamma(\frac{3}{4})}^4} \approx 7.764\,068\,784\ldots,
\end{equation}
in terms of which $\textrm{B}(\frac{1}{4},\frac{1}{2}) = \pi\,\rho^{1/4}$ and
$\textrm{B}(\frac{3}{4},\frac{1}{2}) = 4\,\rho^{-1/4}$.
Inserting (\ref{pe}-\ref{po}) and (\ref{Ie}-\ref{Io}) into (\ref{exactquantization}) gives the
quantization condition
\begin{align}
    \frac{\pi}{3\epsilon} \rho^{1/4}\,E^{3/4} &-
    \frac{\epsilon}{4} \rho^{-1/4}\,E^{-3/4}
    + \sum_{\ell=1}^{\infty}  q^{\text{(e)}}_{\ell}\,\epsilon^{4\ell-1} E^{-(12\ell-3 )/4}  
    \nonumber\\ &+  \sum_{\ell=1}^\infty  q^{\text{(o)}}_{\ell} \epsilon^{4\ell+1}\,E^{-(12\ell+3)/4}
    = \left(N+{\textstyle \frac{1}{2}}\right)\pi.   \label{InfiniteOrderQuantization}
\end{align} 
Here
\begin{align}
    q^{\text{(e)}}_{\ell}  &=  (-1)^{\ell}\,\frac{1}{2}   p_{\ell}^{\text{(e)}}\,
    \frac{(\frac{1}{4})(\frac{1}{4}+1)\cdots(\frac{1}{4}+6\ell-2)}{
    (\frac{1}{2})(\frac{1}{2}+1)\cdots(\frac{1}{2}+6\ell-2)}\,\pi\,\rho^{1/4},\\
    q^{\text{(o)}}_{\ell}  &=  (-1)^{\ell}\,2 p_{\ell}^{\text{(o)}}\,
    \frac{(-\frac{1}{4})(-\frac{1}{4}+1)\cdots(-\frac{1}{4}+6\ell+1)}{
    (\frac{1}{2})(\frac{1}{2}+1)\cdots(\frac{1}{2}+6\ell+1)}\,\rho^{-1/4}.
\end{align}

\begin{figure}[H]
\begin{center}
\includegraphics[clip, trim= 10ex 6.5ex 10ex 5ex,width=0.95\textwidth]{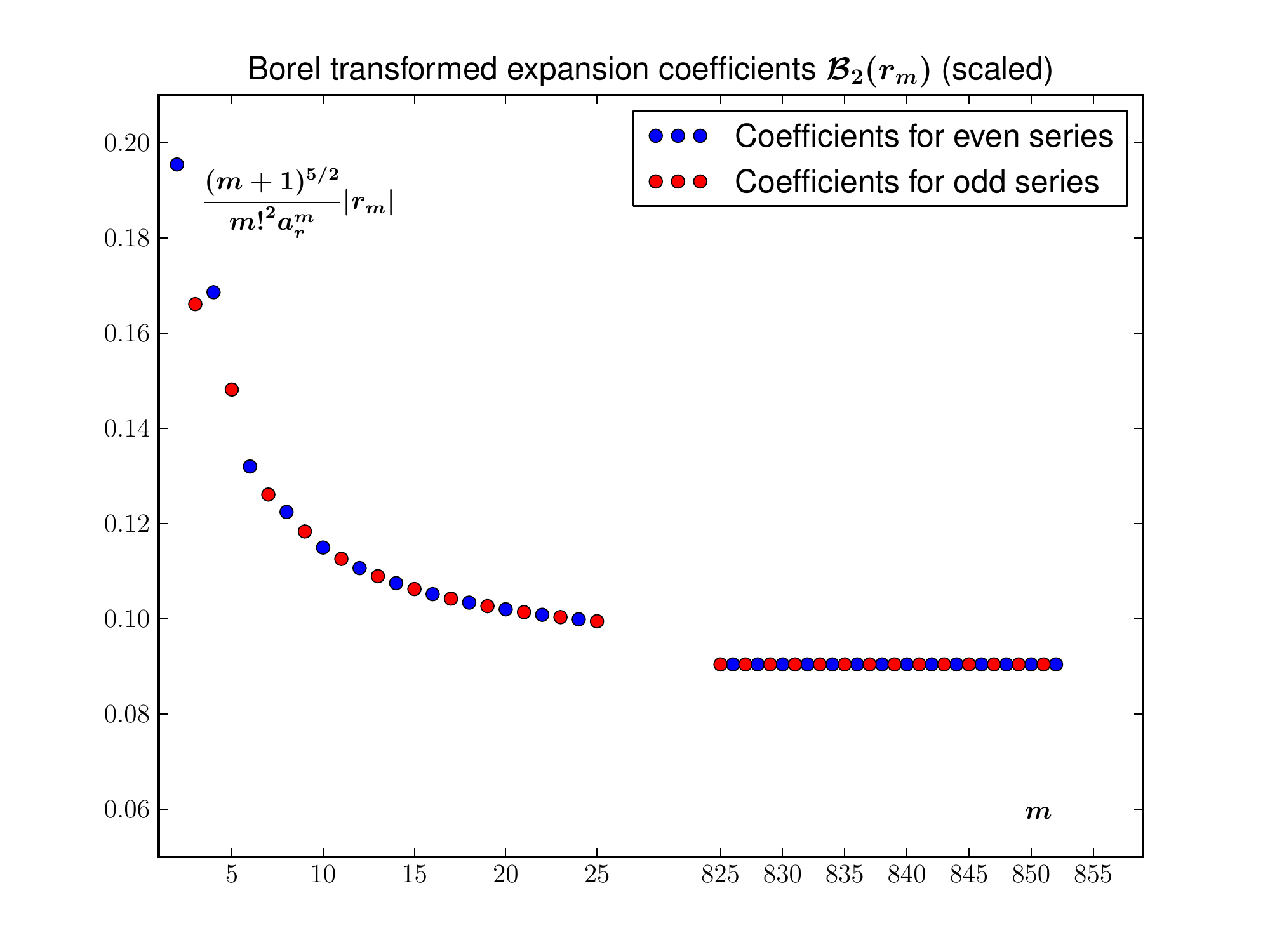}
\end{center}
\caption{Rescaled form of the WKB expansion coefficients ${r}_m$ of
equation (\ref{InfiniteOrderQuantization2}) for $m=0,\ldots,852$.
The index $\nu=\frac{5}{2}$ is chosen as the simplest rational number close to
the best fit,  after which we find $a_r \approx 0.202\,641\,423\,4$ as the best fit to a sequence approaching
a constant absolute value for large $m$.
\label{BRCoeffs}
}
\end{figure}

By introducing
\begin{equation}
  \varepsilon \equiv \kappa^{-1} \epsilon^2\,E^{-3/2},
  \label{RescaledEnergy}
\end{equation}
with $\kappa = \frac{1}{9}\,\rho^{1/2} \approx 0.309\,600\,873\ldots$,
we can rewrite equation (\ref{InfiniteOrderQuantization}) as
\begin{equation}
     \varepsilon  = \frac{1}{(N+\frac{1}{2})^2}\left( 1 + \sum^\infty_{m=1}  r_m\, \varepsilon^m \right)^2
     \label{InfiniteOrderQuantization2}
\end{equation}
with new coefficients $r_{2\ell} = \kappa^{2\ell}\,q^{\text{(e)}}_{\ell}$
and $r_{2\ell+1} = \kappa^{2\ell+1}\,q^{\text{(o)}}_{\ell}$. The first few terms are
\begin{equation*}
  r_1 = -\frac{1}{12\pi},\quad
  r_2 = \frac{11}{41\,472}\,\rho,\quad
  r_3 = \frac{4\,697}{7\,464\,960}\,\frac{\rho}{\pi},\quad
  r_4 = -\frac{390\,065}{8\,026\,324\,992}\rho^2.
\end{equation*}
The further coefficients have the form
\begin{equation*}
  r_{2\ell} = (-1)^{\ell+1}\,\bar{r}_{2\ell}\,\rho^{\ell},\quad
  r_{2\ell+1} = (-1)^{\ell+1}\, \bar{r}_{2\ell+1}\,\rho^{\ell}/\pi,
\end{equation*}
where $\bar{r}_m$ are positive rational numbers. The coefficients $r_m$ grow like $m!^2 a_r^m/(m+1)^\nu$ in magnitude for large $m$.
We have computed these coefficients up to $m=852$ (corresponding to the $1\,704^{\text{th}}$ order of the WKB expansion).
Empirically they fit the cited behaviour quite well, with  $a_r \approx 0.202\,641\,423\,4$ and $\nu = \frac{5}{2}$, 
see figure~\ref{BRCoeffs}.

%\begin{figure}[H]
%\begin{center}
%\includegraphics[clip, trim= 10ex 6.5ex 10ex 5ex,width=0.95\textwidth]{BRCoeffs}
%\end{center}
%\caption{Rescaled form of the WKB expansion coefficients ${r}_m$ of
%equation (\ref{InfiniteOrderQuantization2}) for $m=0,\ldots,852$.
%The index $\nu=\frac{5}{2}$ is choosen as the simplest rational number close to
%the best fit,  after which we find $a_r \approx 0.202\,641\,423\,4$ as the best fit to a sequence approaching
%a constant absolute value for large $m$.
%\label{BRCoeffs}
%}
%\end{figure}

Obviously the sum $r(\varepsilon) \equiv \sum_{m=0}^\infty r_m\,\varepsilon^m$
in equation~(\ref{InfiniteOrderQuantization2}) has zero radius of convergence. However,
if one uses the integral formula
\begin{align}
    m!^2 &=\alpha^{2(m+1)}
    \int_0^{\infty} dx\,x^m\,\text{e}^{-\alpha x}\,
    \int_0^{\infty} dy\,y^m\,\text{e}^{-\alpha y}\nonumber\\
    &=2\alpha^{2(m+1)}\int_0^\infty d\xi\,\xi^m\,K_0(2\alpha\sqrt{\xi}),\quad
    \text{with $\alpha = \text{e}^{\text{i}\phi}$  (${\textstyle -\frac{\pi}{2}< \phi < \frac{\pi}{2}}$),}
\end{align}
and interchange summation and integration, one obtains an integral expression (Borel resummation),
\begin{equation}
   r(\varepsilon) = \int_0^\infty dx\,\text{e}^{-\alpha x}\,\int_0^\infty dy\,\text{e}^{-\alpha y}\,
   \sum_{m=0}^\infty\,\tilde{r}_{m} \left(x y a_r \varepsilon \right)^{m},
   \label{BorelResummation}
\end{equation}
with 
\begin{equation}
\tilde{r}_m = \frac{\alpha^{2(m+1)}}{{m!}^2\,a_r^m}\,r_m.
\end{equation}
Now the sum
\begin{equation}
  \tilde{r}(z) \equiv \sum_{m=0}^\infty \tilde{r}_m\, z^m
  \label{TaylorSeriesIn_z}
\end{equation}
converges for $\vert z \vert < 1$.
For $\alpha=1$ the function $\tilde{r}(z)$ has singularities where $z^2=-1$, 
with the singular parts behaving like $(1+z^2)^{3/2}$ near the singularities.
In terms of the variable ${z^2}/{(1+z^2)}$ this singularity is mapped to $\infty$, 
and the full integration range is mapped to the interval $[0,1]$. However, when one
tries this substitution, in the hope that the (rewritten) sums for $\tilde{r}(z)$  will converge over the
full integration range, one discovers that there are additional singularites
where $z^2 \approx 4$. Hence, to avoid integrating through a singularity, 
one must introduce the phase $\alpha$ (or equivalently integrate along a
different direction in the complex plane). A convenient choice is $\alpha=\text{e}^{\text{i}\pi/8}$,
or its complex conjugate. Actually, to assure a real result after analytic continuation of 
$\tilde{r}(z)$ beyond the radius of convergence of the sum~(\ref{TaylorSeriesIn_z}),
one must take the average of these two choices.
This amounts to taking the real part of the integral~(\ref{BorelResummation}).

\begin{figure}[H]
\begin{center}
\includegraphics[clip, trim= 10ex 6.5ex 10ex 5ex,width=0.95\textwidth]{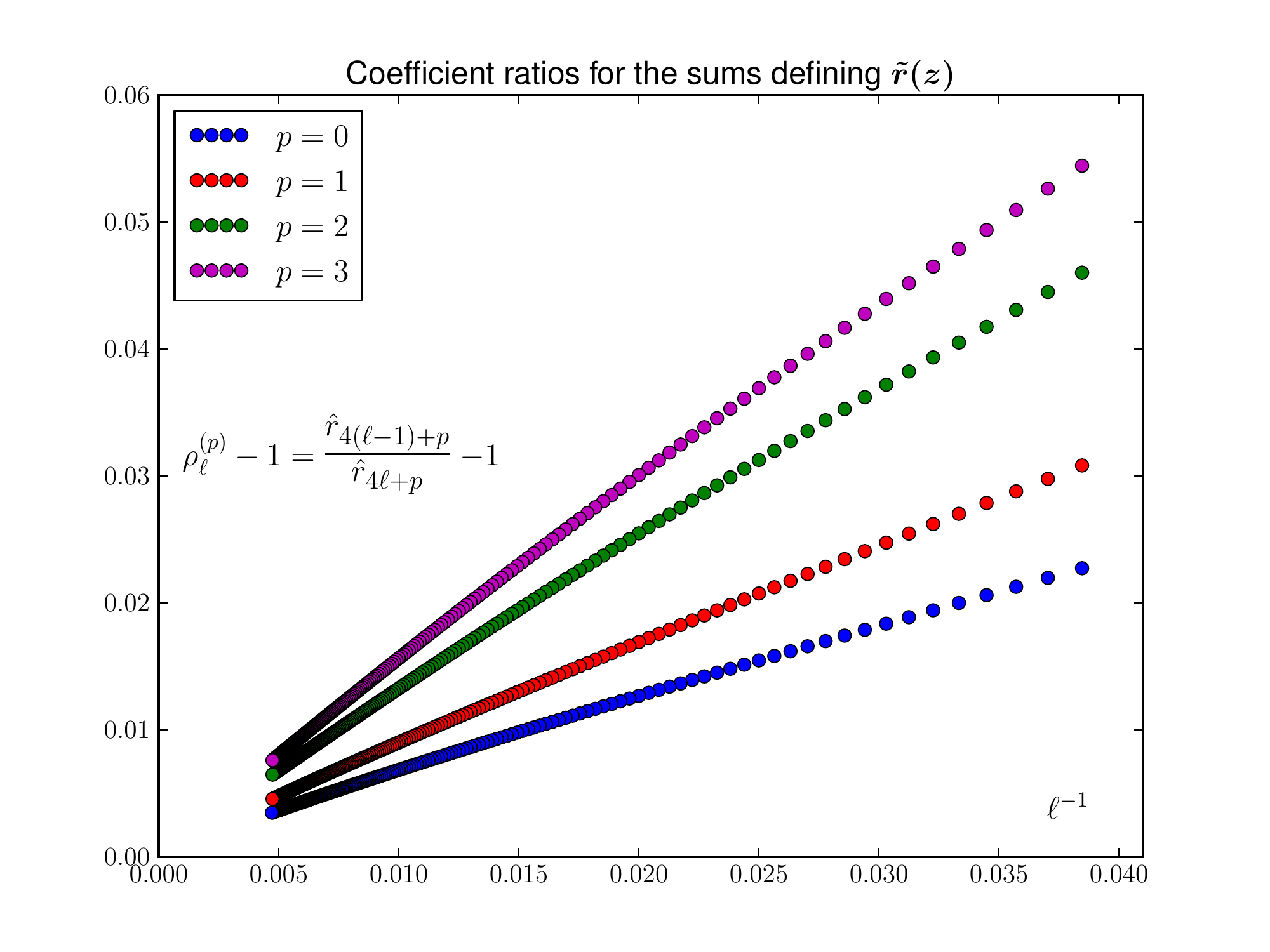}
\end{center}
\caption{\label{BPratios} The convergence radius of the right hand sum in equation~(\ref{PadeResummation})
is, according to the ratio test, given by the inverse ratio $\rho^{(p)}_\ell$ of successive terms in this
sum as $\ell\to\infty$. The computed ratios give convincing evidence that the convergence radius
is unity.
}
\end{figure}

After this choice we separate $\tilde{r}(z)$ into four (infinite) sums,
\begin{equation*}
    \tilde{r}(z) = \sum_{p=0}^3  z^p\, \sum_{\ell=0}^{\infty}  \tilde{r}_{4\ell+p} z^{4\ell}.
\end{equation*}
The function defined by each infinite sum is singular at $z^4=-1$, $z^4\approx -16$, and probably
at infinitely many more points on the negative real $z^4$-axis. Now rewrite
\begin{equation}
      \sum_{\ell\ge0}  \tilde{r}_{4\ell+p}\, z^{4\ell} = \sum_{\ell\ge0} \hat{r}_{4\ell+p}\,\left({\textstyle \frac{z^4}{1+z^4}}\right)^\ell,
      \label{PadeResummation}
\end{equation}
and use the computed coefficients $\tilde{r}_{4\ell+p}$ to find equally many coefficients $\hat{r}_{4\ell+p}$.
By computing the sequence of coefficient ratios 
\begin{equation}
   \rho^{(p)}_{\ell} \equiv \frac{\hat{r}_{4(\ell-1)+p}}{\hat{r}_{4\ell+p}}
\end{equation}
we find empirically (by the ratio test) that the right hand sum of equation (\ref{PadeResummation})
converges for $\vert z^4/(1+z^4) \vert < 1$, see figure~\ref{BPratios}. This completes the construction of a function
\begin{align}
    r(\varepsilon) &= \int_0^{\infty} dx\,\text{e}^{-\alpha x}\,
    \int_0^{\infty} dy\,\text{e}^{-\alpha y}\,\tilde{r}(xya_r\varepsilon)\nonumber\\
    &= 2 \int_0^\infty d\xi\, K_0(2\alpha\sqrt{\xi})\,\tilde{r}(\xi a_r\varepsilon)
\end{align}
which has the same asymptotic expansion as the sum in (\ref{InfiniteOrderQuantization2}), and
where the integrand, notably~$\tilde{r}(z)$, is computable to high precision
over the full integration range.

\section{Series solution of the quantization condition}

The quantization condition~(\ref{InfiniteOrderQuantization}) can now be solved
numerically by first initializing $\varepsilon = \left(N+\textstyle{\frac{1}{2}}\right)^{-2}$,
next iterating the recursion
\begin{equation}
    \varepsilon \leftarrow \left(N+{\textstyle \frac{1}{2}} \right)^{-2}\,r(\varepsilon)^2
\end{equation}
until convergence, and finally applying the relation~(\ref{RescaledEnergy}).

One may also proceed analytically by expressing $\varepsilon$ as a series in the (small) quantity $\delta \equiv (N+\frac{1}{2})^{-2}$,
\begin{equation}
  \varepsilon = \delta + \sum^\infty_{m=2}  s_m\,
    \delta^{m}.
\end{equation}
The coefficients $s_m$ can be computed recursively. The first terms are
\begin{equation}
  s_2 = 2 r_1 = -\frac{1}{6\pi},\quad
  s_3 = 2 r_2 + 5 r^2_1 = \frac{11}{20\,736}\rho + \frac{5}{144\,\pi^2}.
\end{equation}
Computation of the exact $s_m$, which are polynomials in $\rho$ and $\pi^{-1}$ with
rational coefficients, becomes too memory- and time-consuming
beyond the first few tens. We have computed the sequence exactly
up to $s_{56}$, and higher $s_m$ with about $3\,800$ decimals accuracy.

When the sequence of $s_m$ is known one may use~(\ref{RescaledEnergy})
 to express $E$ as a series in $\delta$,
\begin{align}
    E \equiv E_N &=   \kappa^{-2/3}\,\epsilon^{4/3}\, \delta^{-2/3}\,
    \Big{(}1 + \sum_{m\ge1} s_{m+1} \delta^m \Big{)}^{-2/3} \nonumber\\
    &=  \kappa^{-2/3}\,\epsilon^{4/3}\,  \delta^{-2/3}
    \,\Big{(} 1 + \sum_{m\ge1}  t_m \,\delta^m \Big{)}
    \label{EigenvalueSeries}\\
    &\equiv \kappa^{-2/3}\,\epsilon^{4/3}\,\delta^{-2/3}\,t(\delta).
\end{align}

% \begin{figure}[H]
% \begin{center}
% \includegraphics[clip, trim= 10ex 6.5ex 10ex 5ex,width=0.95\textwidth]{BtCoeffs}
% \end{center}
% \caption{\label{BtCoeffs}Rescaled form of the expansion coefficients $t_m$ in 
% equation (\ref{EigenvalueSeries}) for $m=0,\ldots,852$.
% The index $\nu=\frac{5}{2}$ is choosen as the simplest rational number close to
% the best fit,  after which we find $a_t \approx 0.202\,641\,423\,4$ as the best fit to a sequence approaching
% a constant absolute value for large $m$. Note the similarity with the $r_m$-sequence.
% }
% \end{figure}

The first few terms are
\begin{align}
   t_1 &=   \frac{1}{9\pi},\nonumber\\ 
   t_2 &= -\frac{5}{648\,\pi^2}- \frac{11}{31\,104}\rho,\nonumber\\[-1.8ex]
   &\\[-1.8ex]
   t_3 &= \frac{11}{8\,748\pi^3} - \frac{341}{466\,560}\frac{\rho}{\pi},\nonumber\\
   t_4 &= -\frac{1\,309}{5\,038\,848\,\pi^4} + 
   \frac{9\,163}{25\,194\,240\,}\frac{\rho}{\pi^2} + \frac{1\,748\,093}{27\,088\,846\,848}\rho^2.
   \nonumber
\end{align}
In this way an expansion of the WKB-solution (\ref{WKBapp}) to
order $2{\cal N}$ in the quantity $\epsilon$
can be used to compute the expansion (\ref{EigenvalueSeries})
to order ${\cal N}$ in the quantity $(N+\frac{1}{2})^{-2}$. 

\begin{figure}[H]
\begin{center}
\includegraphics[clip, trim= 10ex 6.5ex 10ex 5ex,width=0.95\textwidth]{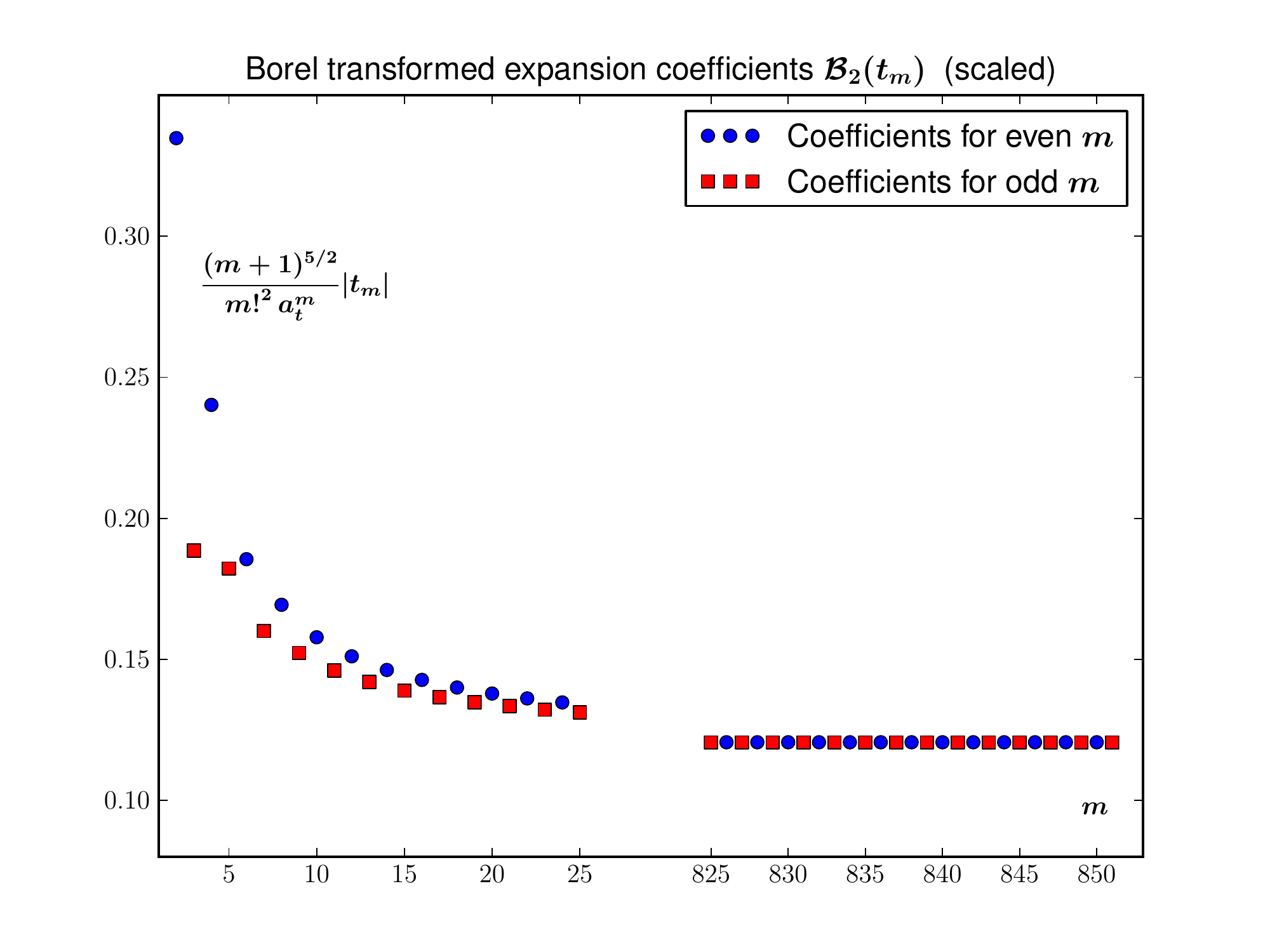}
\end{center}
\caption{\label{BtCoeffs}Rescaled form of the expansion coefficients $t_m$ in 
equation (\ref{EigenvalueSeries}) for $m=0,\ldots,852$.
The index $\nu=\frac{5}{2}$ is chosen as the simplest rational number close to
the best fit,  after which we find $a_t \approx 0.202\,641\,423\,4$ as the best fit to a sequence approaching
a constant absolute value for large $m$. Note the similarity with the $r_m$-sequence.
}
\end{figure}

\section{Extended Borel summation of the asymptotic series}

The sequence of
$t_m$ looks very similar to the sequence of $r_m$, cf. figure~\ref{BtCoeffs}.
Hence one may use the same method to construct a convergent
expression which reproduces the series expansion of $t(\delta)$.

%\begin{figure}[H]
%\begin{center}
%\includegraphics[clip, trim= 8.5ex 6.5ex 10ex 5ex,width=0.95\textwidth]{THcoeffs}
%\end{center}
%\caption{\label{THcoeffs}Scaled versions of the expansion coefficients
%in equation~(\ref{Resummed_Bt-series}). They provide convincing
%evidence for convergence of the sums over the full
%integration range, $0\le z \le \infty$.
%}
%\end{figure}

\begin{figure}[H]
\begin{center}
\includegraphics[clip, trim= 8.5ex 6.5ex 10ex 5ex,width=0.95\textwidth]{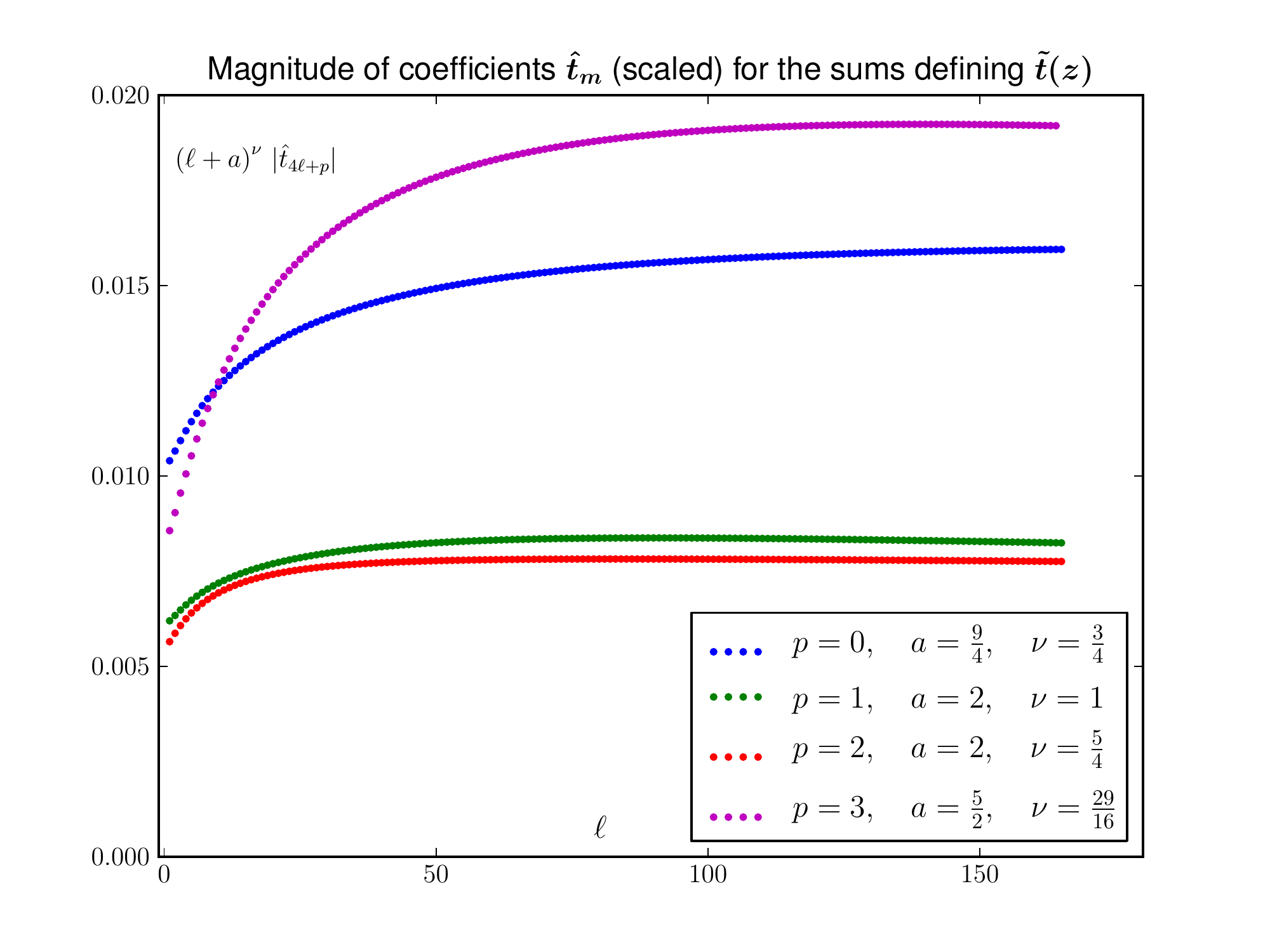}
\end{center}
\caption{\label{THcoeffs}Scaled versions of the expansion coefficients
in equation~(\ref{Resummed_Bt-series}). They provide convincing
evidence for convergence of the sums over the full
integration range, $0\le z \le \infty$.
}
\end{figure}

We define
\begin{equation}
   \tilde{t}_m = \frac{\alpha^{2(m+1)}}{{m!}^2\,a^m_t}\,t_m,
\end{equation}
and coefficients $\hat{t}_m$ such that
\begin{equation}
    \tilde{t}(z) \equiv \sum_{m\ge 0} \tilde{t}_m z^m = 
    \sum_{p=0}^3  z^p \sum_{\ell\ge 0} \hat{t}_{4\ell+p}\,
    \left({\textstyle \frac{z^4}{1+z^4}}\right)^{\ell}.
    \label{Resummed_Bt-series}
\end{equation}
One can use the previously computed coefficients $\tilde{t}_m$ to compute
equally many coefficients $\hat{t}_m$. With the chosen value of
$\alpha = \text{e}^{\text{i}\pi/8}$ the sums over $\ell$ in
equation~(\ref{Resummed_Bt-series}) converge for $0\le z \le \infty$, see figure~\ref{THcoeffs}.

Then, the expansion of the integral expression
\begin{equation}
   t(\delta) = \text{Re}\left\{\int_0^\infty dx\, \text{e}^{-\alpha x}\,\int_0^\infty dy\, 
   \text{e}^{-\alpha y}\, \tilde{t}(xya_t\delta)\right\}
   \label{OriginalBorelIntegral}
\end{equation}
as a series in $\delta$ reproduces the sum in equation~(\ref{EigenvalueSeries}).
However, this particular integral may not be the best way to use the series expansion.

A better approach is to regenerate the series expansion
with an integral expression for the remainder. To this end
write
\begin{equation*}
   \text{e}^{-\alpha x} = -\alpha^* \frac{d}{dx} \text{e}^{-\alpha x}
\end{equation*}
in equation~(\ref{OriginalBorelIntegral}), and perform a partial integration,
\begin{equation*}
     \int_0^\infty\!\! dx\, \text{e}^{-\alpha x}\,\int_0^\infty\!\! dy\,\text{e}^{-\alpha y}\, \tilde{t}(xya_t\delta)
     = t_0 + \alpha^* a_t \delta\, 
     \int_0^\infty\!\! dx\, \text{e}^{-\alpha x}\,\int_0^\infty\!\! dy\,y\,
     \text{e}^{-\alpha y}\, \tilde{t}^{(1)}(xya_t\delta).
\end{equation*}
By repeating this process ${ M}$ times, and taking the real part, one finds
\begin{equation}
     t(\delta) = \sum_{m=0}^{{ M}-1 }t_m\,\delta^m + t^{(M)}_{\text{corr}}(\delta),
     \label{BorelCorrectedExpansion}
\end{equation}
with
\begin{equation}
     t^{(M)}_{\text{corr}}(\delta) = \text{Re}\left\{\left(\alpha^* a_t \delta\right)^{ M} 
     \int_0^\infty\!\! dx\, \text{e}^{-\alpha x}\,\int_0^\infty\!\! dy\,y^{ M}\,
     \text{e}^{-\alpha y}\, \tilde{t}^{({ M})}(xya_t\delta)\right\}.
   \label{BorelCorrection}
\end{equation}
Here 
\begin{align}
  \tilde{t}^{({ M})}(z) &= \frac{d^{ M}}{d z^{ M}} \tilde{t}(z) 
  = \sum_{m\ge 0}   (m+M)(m+M-1)\cdots(m+1)\tilde{t}_{m+{ M}}\,z^m\nonumber\\
  &\equiv \sum_{m\ge 0} \tilde{t}^{(M)}_m\,z^m.
\end{align}
From ${\cal M}$ known coefficients $\tilde{t}_m$ one finds ${\cal M}-M$ coefficients
$\tilde{t}^{(M)}_m$. Again rewrite, cf.~equation~(\ref{Resummed_Bt-series}),
\begin{equation}
    \tilde{t}^{(M)}(z) = \sum_{m\ge 0} \tilde{t}^{(M)}_m\,z^m = 
    \sum_{p=0}^3  z^p \sum_{\ell\ge0}\,\hat{t}^{(M)}_{4\ell + p} 
    \left({\textstyle \frac{z^4}{1+z^4}}\right)^{\ell},
    \label{tM_series}
\end{equation}
and use the known coefficients $\tilde{t}^{(M)}_m$ to compute equally many
coefficients $\hat{t}^{(M)}_m$.

%\begin{figure}[H]
%\begin{center}
%\includegraphics[clip, trim= 10ex 6.5ex 10ex 1.75ex,width=0.95\textwidth]{WKB_behaviour.pdf}
%\end{center}
%\caption{\label{WKB_behaviour}
%Behaviour of the WKB expansion of the few lowest
%eigenvalues $E_N$. The blue points show the (asymptotic) sum $\sum_{m=0}^{M-1} t_m\,\delta^m$
%in equation~(\ref{BorelCorrectedExpansion}). The green line shows the full expression
%$t(\delta)$; it is independent of $M$ to the expected numerical accuracy
%of the correction term (\ref{BorelCorrection}). The red line shows the exact eigenvalue,
%evaluated numerically by our very-high-precision routine. The
%Borel corrected WKB series does not converge towards the exact eigenvalue.
%}
%\end{figure}

\begin{figure}[H]
\begin{center}
\includegraphics[clip, trim= 10ex 6.5ex 10ex 1.75ex,width=0.95\textwidth]{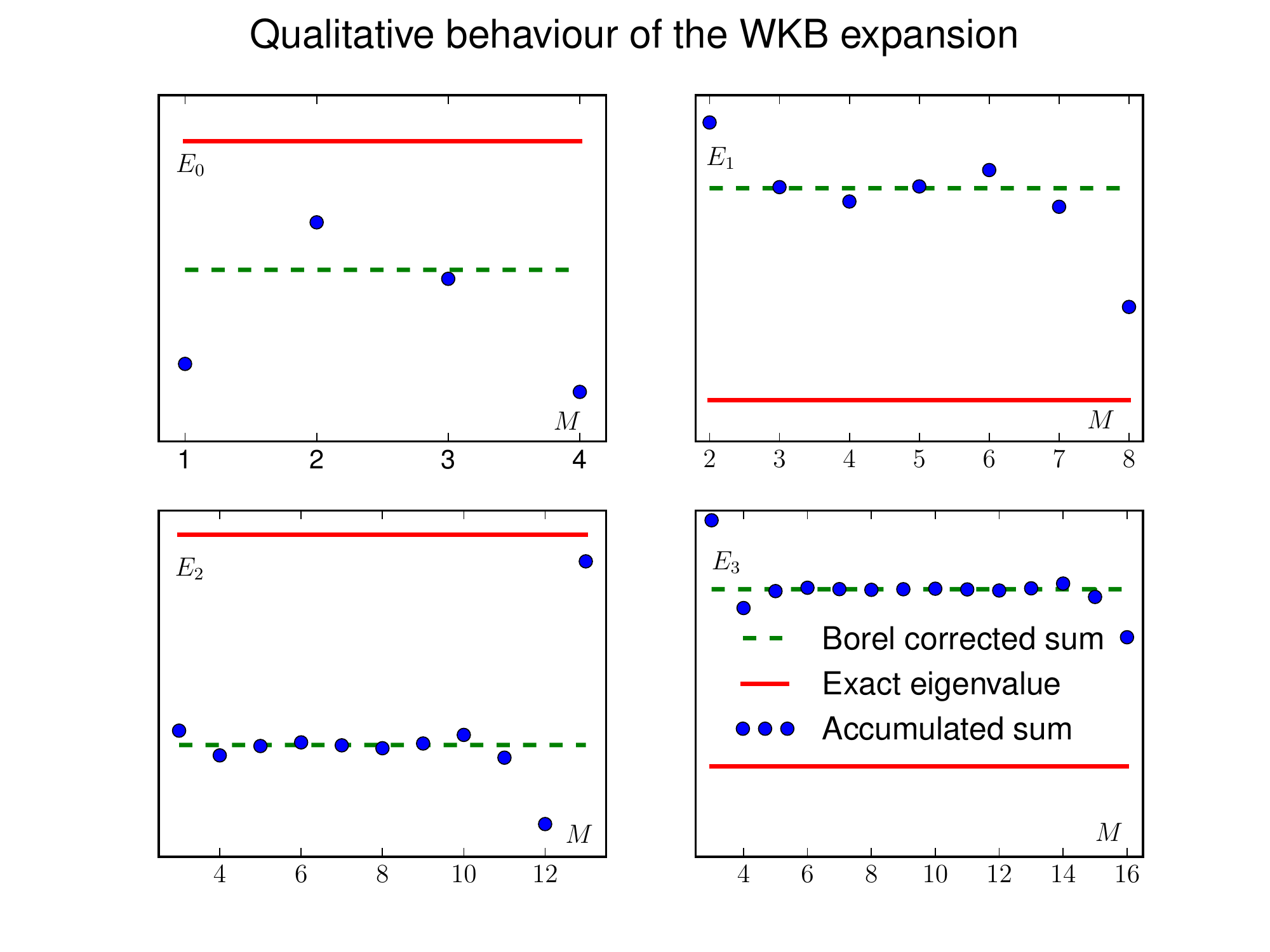}
\end{center}
\caption{\label{WKB_behaviour}
Behaviour of the WKB expansion of the few lowest
eigenvalues $E_N$. The blue points show the (asymptotic) sum $\sum_{m=0}^{M-1} t_m\,\delta^m$
in equation~(\ref{BorelCorrectedExpansion}). The green line shows the full expression
$t(\delta)$; it is independent of $M$ to the expected numerical accuracy
of the correction term (\ref{BorelCorrection}). The red line shows the exact eigenvalue,
evaluated numerically by our very-high-precision routine. The
Borel corrected WKB series does not converge towards the exact eigenvalue.
}
\end{figure}

We finally insert the expansion (\ref{tM_series}) into (\ref{BorelCorrection}) and
perform the integral numerically. This gives $t^{(M)}_{\text{corr}}(\delta)$ to a relative
accuracy of about $10^{-10}$. As a consistency check we verify that $t(\delta)$ is
independent of $M$, at least for $M$-values around the point where
$\big{|} t^{(M)}_{\text{corr}}(\delta) \big{|}$ is minimum. As can be seen qualitatively
from figure~\ref{WKB_behaviour} this works well for the lowest eigenvalues. But it also
shows that the WKB-series does not reproduce the exact eigenvalues, even when the
correction term (\ref{BorelCorrection}) is included. The quantitative results are
shown numerically for the two lowest eigenvalues in tables~\ref{E0}-\ref{E1}.

\begin{figure}[h]
\begin{center}
\includegraphics[clip, trim= 10ex 6.5ex 10ex 1.75ex,width=0.82\textwidth]{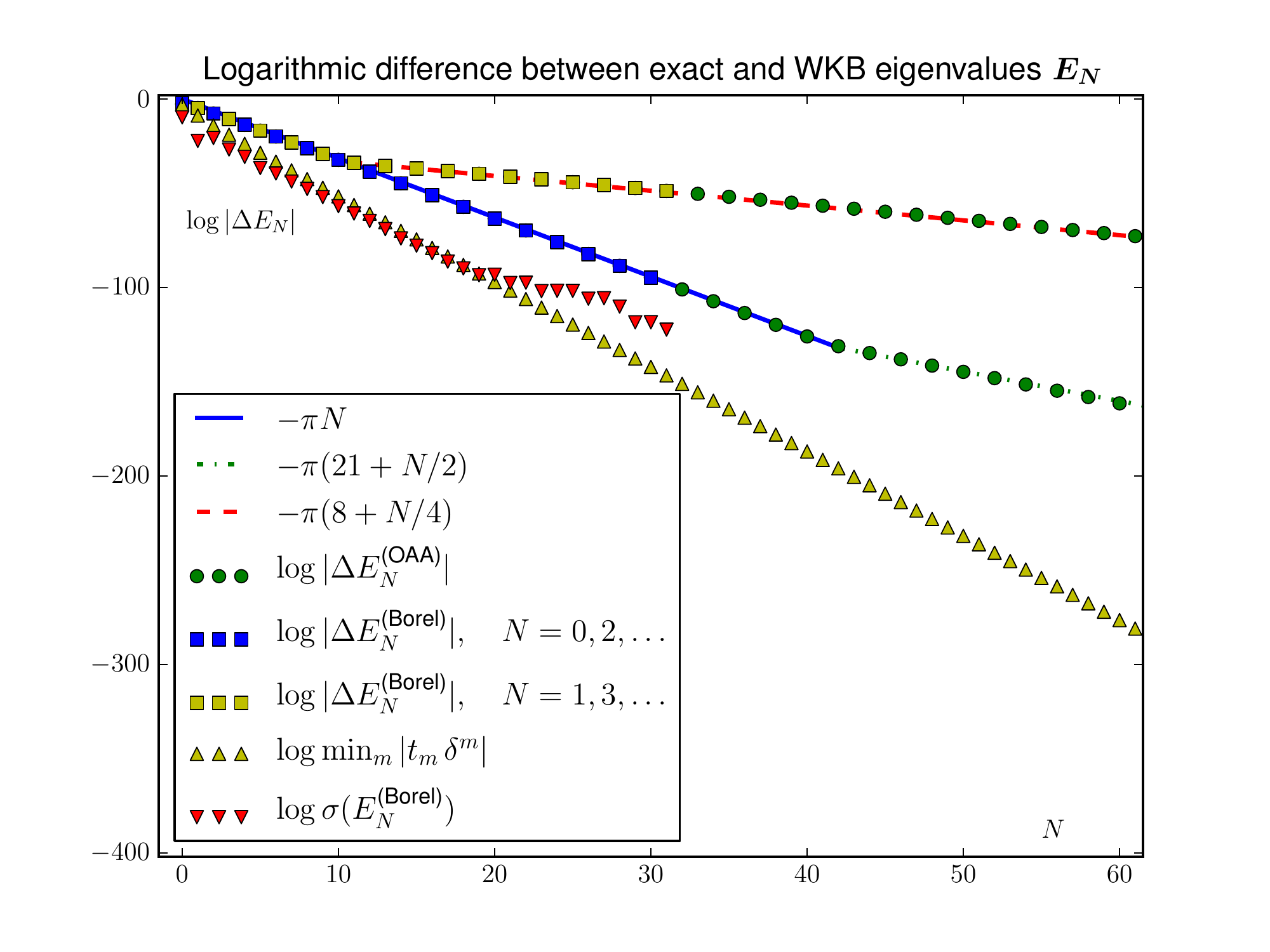}
\end{center}
\caption{\label{logDeltaEn2}Difference between the exact eigenvalues $E_{N,\text{exact}}$
(computed numerically to very high precision) and the WKB eigenvalues $E_{N,\text{WKB}}$,
computed using either the optimal asymptotic approximation (OAA) or adding the correction
integral from Borel resummation (Borel). The results of these two methods cannot be distinguished
in the figure when $N\ge 1$.
The later is found from equation~(\ref{RescaledEnergy}),
with $t(\delta)$ computed from equation~(\ref{BorelCorrectedExpansion}) for a range of $M$-values around
$2.2 N$. The result varies little with $M$, as indicated by the plotted standard deviation $\sigma(E_N)$.
Hence, the difference between $E_{N,\text{exact}}$ and $E_{N,\text{WKB}}$ is much larger than the
uncertainty in $E_{N,\text{WKB}}$ due to numerical evaluation of the integral~(\ref{BorelCorrection}),
although exponentially small as function of $N$. The correction terms look quite simple,
with an interesting difference between the even and odd eigenvalues.
}
\end{figure}

The WKB-series shows a quite stable result when the correction term from Borel
resummation is added, with an uncertainty much smaller than the distance to the exact
result. This can be seen for a larger range of eigenvalues
in figure~\ref{logDeltaEn2}, where we plot $\log\vert E_{N,\text{exact}} - E_{N,\text{WKB}}\vert$
as function of $N$.

We find empirically that 
\begin{equation*}
 E_{N,\text{exact}} > E_{N,\text{WKB}}\quad \text{for all even } N=2M,
\end{equation*}
with a difference which varies like $\text{e}^{-\pi N}$ with
for $N\le 42$, and approximately like $\text{e}^{-\pi(21+ N/2)}$
for $N\ge 42$. Further
\begin{equation*}
  E_{N,\text{exact}} < E_{N,\text{WKB}}\quad\text{for odd } N=2M+1 \le 9,
\end{equation*}
also with a difference which varies like $\text{e}^{-\pi N}$ with $N$,
and
\begin{align*}
    E_{N,\text{exact}} &< E_{N,\text{WKB}}\quad\text{for odd } N=4M+3 \ge 11,\\
    E_{N,\text{exact}} &> E_{N,\text{WKB}}\quad\text{for odd } N=4M+1 \ge 13.
\end{align*}
In these cases the difference behaves approximately
like $\text{e}^{-\pi(8+ N/4)}$.

\begin{table}[H]
\begin{center}
\caption{Behaviour of WKB expansion for $E_0$}\label{E0}
\begin{tabular}{|c|c|c|c|}
\hline
&&&\\[-2.0ex]
\multicolumn{1}{|c|}{$M$}&\multicolumn{1}{c|}{$E^{(M-1)}_{0,\text{WKB}}$}&\multicolumn{1}{c|}{$E^{(M-1)}_{0,\text{WKB}}+E^{(M)}_{0,\text{corr}}$}&\multicolumn{1}{c|}{$E_{0,\text{exact}}$}\\[0.6ex]
\hline
&&&\\[-2.0ex]
1&0.867\,145\,326\,484\,821&0.949\,048\,242\,147\,079&1.060\,362\,090\,484\,183\\
2&0.989\,821\,295\,452\,906&0.949\,048\,242\,213\,528&1.060\,362\,090\,484\,183\\
3&0.940\,878\,506\,803\,713&0.949\,048\,245\,949\,142&1.060\,362\,090\,484\,183\\
4&0.842\,885\,181\,871\,221&0.949\,048\,880\,595\,005&1.060\,362\,090\,484\,183\\
\hline
\end{tabular}
\caption*{Behaviour of the WKB series for the lowest eigenvalue $E_0$.
The first column shows the results of summing the first $M$ terms of
the WKB-series. The second column the result after the correction term
(\ref{BorelCorrection}) from Borel resummation has been added. The last
column shows the eigenvalue computed numerically to very high precision.  
}
\end{center}
\end{table}

\begin{table}[H]
\begin{center}
\caption{Behaviour of WKB expansion for $E_1$}\label{E1}
\begin{tabular}{|c|c|c|c|}
\hline
&&&\\[-2.0ex]
\multicolumn{1}{|c|}{$M$}&\multicolumn{1}{c|}{$E^{(M-1)}_{1,\text{WKB}}$}&\multicolumn{1}{c|}{$E^{(M-1)}_{1,\text{WKB}}+E^{(M)}_{1,\text{corr}}$}&\multicolumn{1}{c|}{$E_{1,\text{exact}}$}\\[0.6ex]
\hline
&&&\\[-2.0ex]
1&3.751\,919\,923\,550\,433&3.808\,235\,541\,533\,203&3.799\,673\,029\,801\,394\\
2&3.810\,896\,378\,060\,855&3.808\,235\,541\,533\,340&3.799\,673\,029\,801\,394\\
3&3.808\,282\,018\,212\,746&3.808\,235\,541\,531\,506&3.799\,673\,029\,801\,394\\
4&3.807\,700\,409\,855\,639&3.808\,235\,541\,531\,468&3.799\,673\,029\,801\,394\\
5&3.808\,311\,380\,649\,850&3.808\,235\,541\,532\,831&3.799\,673\,029\,801\,394\\
6&3.808\,972\,737\,814\,702&3.808\,235\,541\,513\,511&3.799\,673\,029\,801\,394\\
7&3.807\,487\,485\,686\,370&3.808\,235\,542\,141\,864&3.799\,673\,029\,801\,394\\
8&3.803\,436\,692\,708\,719&3.808\,235\,536\,164\,707&3.799\,673\,029\,801\,394\\
\hline
\end{tabular}
\caption*{Behaviour of the WKB series for the eigenvalue $E_1$.
The first column shows the results of summing the first $M$ terms of
the WKB-series. The second column the result after the correction term
(\ref{BorelCorrection}) from Borel resummation has been added. The last
column shows the eigenvalue computed numerically to very high precision.  
}
\end{center}
\end{table}

It should be clear that the Dunham quantization formula~(\ref{exactquantization})
does not provide exact eigenvalues in this case. There are (leading) order correction
terms which look intriguingly simple. They are manifestations of the fact that the
WKB approximation is inexact, even when summed to arbitrarily high order. Starting
with a WKB-solution which behaves like
\begin{equation*}
     Q(z)^{-1/4}\, \text{exp}\left(-\frac{1}{\epsilon}\int^z_{z_0} \text{d}t \sqrt{-Q(t)} + \cdots \right),
\end{equation*}
no higher-order correction will provide a contribution which changes the sign of
the square root, i.e., provides a solution which behave like
\begin{equation*}
     Q(z)^{-1/4}\, \text{exp}\left(\frac{1}{\epsilon}\int^z_{z_0} \text{d}t \sqrt{-Q(t)} + \cdots \right).
\end{equation*}
However, both behaviours are usually present in the exact solution. In asymptotic analysis they are
said to emerge when Stokes lines are crossed.

%
% THE LAST STUFF: BIBLIOGRAPHY, PAPERS
%
%\backmatter

% Bibliography

\bibliographystyle{JHEP} % Using JHEP style bibliography as an example
%\bibliography{phd_template_bibliography}

\begin{thebibliography}{00}
\refstepcounter{chapter}\addcontentsline{toc}{chapter}{\hspace{4.3ex}Bibliography}

%1
\bibitem{ShigeruKondo}
Alexander J.~Yee, Shigeru ~Kondo~\textit{et. al.}, \texttt{http://www.numberworld.org/\\misc\_runs/pi-10t/details.html} (2011)

%2
\bibitem{NobelPrize2005}
The Royal Swedish Academy of Sciences, \texttt{http://www.nobelprize.org/\\nobel\_prizes/physics/laureates/2005/\\advanced-physicsprize2005.pdf}

% \bibitem{Kanada_etal}
%   Y.~Kanada {\em et. al.\/},
%   {\tt http://www.super-computing.org/$\backslash$\\pi\_current.html\/} (2005)
% 
% \bibitem{New_World_Record}
% Alexander J.~Yee, Shigeru ~Kondo~{\em et. al.\/}, {\tt http://www.numberworld.org/misc\_runs/pi-5t/announce_en.html\/} (2010)

%3
\bibitem{PDG}
J.~Beringer~\textit{et. al.} (Particle Data Group) \textit{The Review of Particle Physics},
Phys.~Rev.~D~\textbf{86}, 010001 (2012)

%4
\bibitem{Electron_g-2_10thOrder}
T.~Aoyama, M.~Hayakawa, T.~Kinoshita, and M.~Nio,
\textit{Tenth-Order QED Contribution to the Electron ${g-2}$ and an Improved Value of the Fine Structure Constant},
Phys. Rev. Lett. \textbf{109}, 111807 (2012)

%5
\bibitem{Muon_g-2_10thOrder}
T.~Aoyama, M.~Hayakawa, T.~Kinoshita, and M.~Nio,
\textit{Complete Tenth-Order QED contribution to the Muon ${g-2}$},
Phys. Rev. Lett. \textbf{109}, 111808 (2012)

%6
\bibitem{Fuchs}
L.~Fuchs,
\textit{Zur Theorie der linearen Differentialgleichungen mit ver{\"a}nderlichen Coefficienten},
Journal f{\"u}r die reine und angewandte Mathematik \textbf{66}, 121 (1866)

%7
\bibitem{FrobeniusMethod}
F.G.~Frobenius,
\textit{\"Uber die Integration der linearen Differentialgleichungen durch Reihen},
Journal f{\"u}r die reine und angewandte Mathematik,
\textbf{76}, 214 (1873)

%8
\bibitem{MorseFeshbach}
Philip~M.~Morse and Herman~Feshbach,
\textit{Methods of Theoretical Physics, Part I}, 508--515, 
McGraw-Hill (1953)

%9
\bibitem{MorseFeshbach2}
Philip~M.~Morse and Herman~Feshbach,
\textit{Methods of Theoretical Physics, Part I}, 663, 
McGraw-Hill (1953)

%10
\bibitem{Fokker}
A.D.~Fokker,
\textit{Die mittlere Energie rotierender elektrischer Dipole im Strahlungsfeld}, 348,
Ann. Phys. 810--820 (1914)

%11
\bibitem{MaxPlanck}
M.~Planck,
\textit{Sitz.ber} (1917)

%12
\bibitem{CLN} B.~Haible and R.B.~Kreckel,
\textit{CLN -- Class Library for Numbers},
\texttt{http://www.ginac.de/CLN/}

%13
\bibitem{GMP} T.~Granlund and collaborators,
\textit{GMP -- The GNU Multiple Precision Arithmetic Library},
\texttt{http://gmplib.org/}

%14
\bibitem{SchonhageStrassen}
A.~Sch{\"o}nhage and V.~Strassen,
\textit{Schnelle Multiplikasjon gro{\ss}er Zahlen},
Computing~\textbf{7}, 281--292 (1971)

% \bibitem{PDG}
% K.~Nakamura~{\em et. al.\/} {\em The Review of Particle Physics\/},
% J.~Phys.~G~{\bf 37}, 075021 (2010)

%14a
\bibitem{Jentschura_ZinnJustinI}
J.~Zinn-Justin and U.D.~Jentschura,
{Multi-Instantons and Exact Results I: Conjectures, WKB Expansions, and Instanton Interactions},
Annals of Physics \textbf{313}, 197--267 (2004)
\texttt{arXiv:quant-ph/0501136}

%14b
\bibitem{AsifAnneKare}
A.~Mushtaq, A.~Kv{\ae}rn{\"o} and K.~Olaussen,
\textit{Systematic Improvements of Splitting Methods for the Hamilton Equations},
Proceedings of The World Congress on Engineering 2012 Vol I, WCE 2012, July 4-6, 2012, 
London, U.K., 247--251.
\texttt{arXiv:1204.4117}

%15
\bibitem{NielsBohrAug1913}
Niels Bohr,
\textit{On the Constitution of Atoms and Molecules},
Philosophical Magazine Series 6, \textbf{26}, 1--25 (1913)

%16
\bibitem{NielsBohrSep1913}
Niels Bohr,
\textit{On the Constitution of Atoms and Molecules} Part II,
Philosophical Magazine Series 6, \textbf{26}, 476--502 (1913)

%17
\bibitem{Wilson1915}
William~Wilson,
\textit{The quantum theory of radiation and line spectra},
Philosophical Magazine Series 6, \textbf{29}, 795--802 (1915) 

%18
\bibitem{Sommerfeld1916}
Arnold~Sommerfeld,
\textit{Zur Quantentheorie der Spectrallinien},
Annalen der Physik \textbf{51}, 1 (1916)

%19
\bibitem{Jeffreys}
H.~Jeffreys,
\textit{On certain approximate solutions of linear differential equations of the second order},
Proceedings of the London Mathematical Society \textbf{23}, 428--436 (1924)

%20
\bibitem{Wentzel}
G.~Wenzel,
\textit{Eine Verallgemeinerung der Quantenbedingungen für die Zwecke der Wellenmechanik},
Zeitschrift f{\"u}r Physik \textbf{38}, 518--529 (1926)

%21
\bibitem{Kramers}
H.A.~Kramers,
\textit{Wellenmechanik und halbz{\"a}hlige Quantisierung},
Zeitschrift f{\"u}r Physik \textbf{39}, 828--840 (1926)

%22
\bibitem{Brillouin}
L.~Brillouin,
\textit{La m{\'e}canique ondulatoire de Schr{\"o}dinger:
une m{\'e}thode g{\'e}n{\'e}rale de resolution par approximations successives},
Comptes Rendus de l'Academie des Sciences \textbf{183}, 24--26 (1926)

%23
\bibitem{Schiff}
L.I.~Schiff, 
\textit{Quantum Mechanics}, Third Edition, section 34, McGraw-Hill (1968) 

%24
\bibitem{Kroemer}
H.~Kroemer,
\textit{Quantum Mechanics: for engineering, materials science, and applied physics},
Chapter 6, Prentice Hall (1994)                                            

%25
\bibitem{BenderOrszag}
C.M~Bender and S.A.~Orszag, 
\textit{Advanced Mathematical Methods for Scientists and Engineers},
Chapter 10, McGraw-Hill (1978)

%26
\bibitem{Dunham}
J.L.~Dunham, 
\textit{The Wentzel-Brillouin-Kramers Method of Solving the Wave Equation\/},
Phys.~Rev.~\textbf{41}, 713 (1932)

%27
\bibitem{Bender_etal}
C.M.~Bender, K.~Olaussen, and P.S.~Wang,
\textit{Numerological analysis of the WKB approximation in large order},
Physical Review \textbf{D16}, 1740--1748 (1977)

%28
\bibitem{Bailey1964}
Paul~B.~Bailey,
\textit{Exact Quantizatization Rules for the OneDimensional
Schr{\"o}dinger Equation with Turning Points},
Journal of Mathematical Physics \textbf{5}, 1293--1297 (1964)

%29
\bibitem{RosenzweigKrieger}
C.~Rosenzweig and J.B.~Krieger,
\textit{Exact Quantization Conditions},
Journal of Mathematical Physics~\textbf{9}, 849--860 (1968)

%30
\bibitem{MorsePotential}
Philip~M.~Morse,
\textit{Diatomic molecules according to the wave mechanics. II. Vibrational levels},
Phys.~Rev.~\textbf{34}, 57--64 (1929)

%31
\bibitem{PoschlTellerPotential}
G.~P{\"o}schl and E.~Teller,
\textit{Bemerkungen zur Quantenmechanik des anharmonischen Ozillators},
Zeitschrift f{\"u}r Physik \textbf{83}, 143--151 (1933)

%32
\bibitem{SergeLang}
Serge Lang,
\textit{Linear Algebra} Third Edition, 262--264,
Springer-Verlag (1987) 

\bibitem{HighPrecisionSolutions}
Asif Mushtaq, Amna Noreen, K{\aa}re Olaussen, Ingjald {\O}verb{\o},
\textit{Very-high-precision solutions of a class of Schr{\"o}dinger type equations},
Computer Physics Communications \textbf{182}, 1810--1813 (2011);
\texttt{arXiv:1008.0834}

\bibitem{NormalizedEigenfunctions}
Amna Noreen and K{\aa}re Olaussen,
\textit{Very-high-precision normalized eigenfunctions for
a class of Schr{\"o}dinger type equations},
Proceedings of World Academy of Science, Engineering and Technology \textbf{76}, 831--836 (2011);
\texttt{arXiv:1105.1460}

\bibitem{SeriesSolutions}
Amna Noreen and K{\aa}re Olaussen,
\textit{High precision series solution of differential equations:
Ordinary and regular singular point of second order ODEs},
Computer Physics Communications \textbf{183}, 2291--2297 (2011);
\texttt{arXiv:1205.2226}

\bibitem{WKBLegendre}
Amna Noreen and K{\aa}re Olaussen,
\textit{Estimating Coefficients of Frobenius Series by
Legendre Transform and WKB Approximation}
Proceedings of the World Congress of Engineering 2012 Vol II WCE 2012, London U.K. 789--791;
\texttt{arXiv:1205.2221}


\bibitem{GeneratingFrobeniusSeries}
Amna Noreen and K{\aa}re Olaussen,
\textit{Generating Very-High-Precision Frobenius Series
with Apriori Estimates of Coefficients},
IAENG International Journal of Computer Science \textbf{39}, 386-393 (2012);
\texttt{arXiv:1209.6237}

\bibitem{LoopExpansion}
Amna Noreen and K{\aa}re Olaussen,
\textit{Quantum loop expansion to high orders, extended Borel summation, 
and comparison with exact results},
\texttt{arXiv:1209.6242}

\end{thebibliography}

\end{document}